\documentclass[acmnow]{acmtrans2m}

\newdef{definition}[theorem]{Definition}
\newdef{remark}[theorem]{Remark}

\usepackage{graphics}
\usepackage{rotating}
\usepackage{verbatim}
\usepackage{longtable}
\usepackage{lscape}
\usepackage[tight,footnotesize]{subfigure}

\title{When Things Matter: A Data-Centric View of the Internet of Things}

\author{YONGRUI QIN, The University of Adelaide\\
QUAN Z. SHENG, The University of Adelaide\\
NICKOLAS J.G. FALKNER, The University of Adelaide\\
SCHAHRAM DUSTDAR, Vienna University of Technology\\
HUA WANG, Victoria University\\
ATHANASIOS V. VASILAKOS, National Technical University of Athens
}

\begin{abstract}
With the recent advances in radio-frequency identification (RFID), low-cost wireless sensor devices, and Web technologies, the Internet of Things (IoT) approach has gained momentum in connecting everyday objects to the Internet and facilitating machine-to-human and machine-to-machine communication with the physical world. While IoT offers the capability to connect and integrate both digital and physical entities, enabling a whole new class of applications and services, several significant challenges need to be addressed before these applications and services can be fully realized. A fundamental challenge centers around managing IoT data, typically produced in dynamic and volatile environments, which is not only extremely large in scale and volume, but also noisy, and continuous. This article surveys the main techniques and state-of-the-art research efforts in IoT from data-centric perspectives, including data stream processing, data storage models, complex event processing, and searching in IoT. Open research issues for IoT data management are also discussed.

\end{abstract}

\category{C.2.4}{Distributed Systems}{Distributed Databases}
\category{H.3.5}{Information Storage and Retrieval}{On-line Information Services}
[Web-based services]

\terms{Algorithms, Performance}

\keywords{Internet of Things, data management, applications}

\begin{document}

\begin{bottomstuff}
Authors' addresses: Y. Qin, Q. Z. Sheng, N. Falkner, School of Computer Science, The University of Adelaide, Adelaide SA 5005, 
Australia; emails: {\tt\{yongrui, qsheng, nick\}@cs.adelaide.edu.au}. S. Dustdar, Institute of Information Systems, Vienna University of Technology; email: 
{\tt dustdar@infosys.tuwien.ac.at}. H. Wang, Center for Applied Informatics, Victoria University, Melbourne, 
Victoria 3122, Australia; email: {\tt hua.wang@vu.edu.au}; A. Vasilakos, Department of Electrical and Computer Engineering, National Technical University of Athens, Greece; email: {\tt vasilako@ath.forthnet.gr}.

\end{bottomstuff}

\maketitle

\section{Introduction}\label{sec:intro}

The Internet is a global system of networks interconnecting computers using the standard Internet protocol suite. It has significant impact on the world as it can serve billions of users worldwide. Millions of private, public, academic, business, and government networks, of local to global scope, all contribute to the formation of the Internet. It is a network of networks and each network connects various numbers of computers. Hence, the traditional Internet has a focus on computers and can be called the Internet of Computers. In contrast, evolving from the Internet of Computers, the Internet of Things (IoT) emphasizes things rather than computers \cite{AshtonK09}. It aims to connect everyday objects, such as coats, shoes, watches, ovens, washing machines, bikes, cars, even humans, plants, animals, and changing environments, to the Internet to enable communication/interactions between these objects. The ultimate goal of IoT is to enable computers to see, hear and sense the real world. It is predicted by Ericsson that the number of Internet-connected things will reach 50 billion by 2020. Electronic devices and systems exist around us providing different services to the people in different situations: at home, at work, in their office, or driving a car on the street~\cite{JamesCJS09}.

``Changes brought about by the Internet will be dwarfed by those prompted by the networking of everyday objects'', says a report by a United Nation (UN) \cite{Biddlecombe05}. IoT is widely regarded as the number one of top 10 technologies that will change the world in the next 10 years \cite{Bort11}. National Intelligence Council \cite{NICDCT08} foresees that ``by 2025, Internet nodes may reside in everyday things $-$ food packages, furniture, paper documents, and more. Widespread diffusion of an Internet of Things (IoT) could contribute invaluably to economic development."

\begin{figure}[t]
\centering
\scalebox{0.67}{\includegraphics{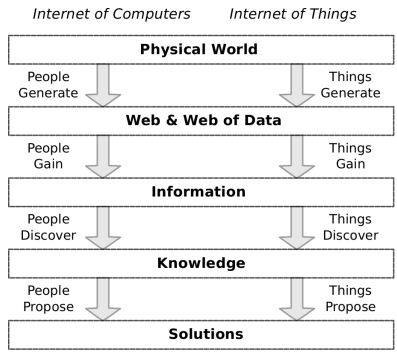}}
\caption{Internet of Computers v.s. Internet of Things}
\label{fig_people_and_things}
\end{figure}

\vspace{1mm
}

There are several definitions or visions of IoT from different perspectives. From the viewpoint of services provided by things, IoT means ``a world where things can automatically communicate to computers and each other providing services to the benefit of the human kind" \cite{CASAGRAS}. From the viewpoint of connectivity, IoT means ``from anytime, anyplace connectivity for anyone, we will now have connectivity for anything" \cite{ITU05}. 
From the viewpoint of communication, IoT refers to ``a world-wide network of interconnected objects uniquely addressable, based on standard communication protocols" \cite{INFSO08}. Finally, from the viewpoint of networking, IoT is the Internet evolved ``from a network of interconnected computers to a network of interconnected objects" \cite{EUCommission09}.

We focus on our study of the Internet of Things from a {\em data perspective}. As shown in Fig. \ref{fig_people_and_things}, data is processed differently in the Internet of Things and traditional Internet environments (i.e., Internet of Computers). In the Internet of Computers, both main data producers and consumers are human beings. However, in the Internet of Things, the main actors become {\em things}, which means things are the majority of data producers and consumers. Therefore, we give our definition of the Internet of Things as follows:

``{\em In the context of the Internet, addressable and interconnected things, instead of humans, act as the main data producers, as well as the main data consumers. Computers will be able to learn and gain information and knowledge to solve real world problems directly with the data fed from things. As an ultimate goal, computers enabled by the Internet of Things technologies will be able to sense and react to the real world for humans.}''

\vspace{1mm}
As of 2012, 2.5 quintillion ($2.5 \times 10^{18}$) bytes of data are created daily\footnote{http://www-01.ibm.com/software/data/bigdata/}. In IoT, connecting all of the things that people care about in the world becomes possible. All these things would be able to produce much more data than nowadays \cite{barnagi-is2013}. The volumes of data are vast, the generation speed of data is fast and the data/information space is global \cite{JamesCJS09}. Indeed, IoT is one of the major driving forces for {\em big data analytics}. 
Given the scale of IoT, topics such as storage, distributed processing, real-time data stream analytics, and event processing are all critical, and we may need to revisit these areas to improve upon existing technologies for applications of this scale \cite{barnagi-is2013,JamesCJS09}. 

\begin{figure}[t]
\centering
\scalebox{0.67}{\includegraphics{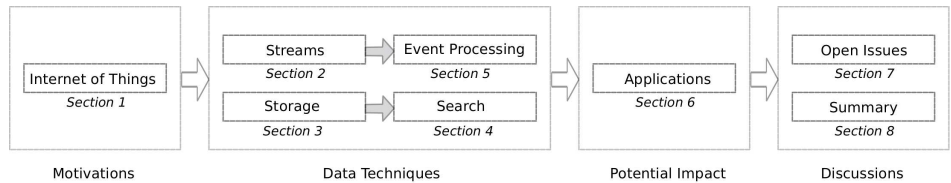}}
\caption{Roadmap of this survey}
\label{fig_survey_roadmap}
\end{figure}

In this survey, we systematically investigate the key technologies related to the development of IoT and its applications, particularly from a data-centric perspective. The aim of this work is to provide a better understanding of the current research activities and issues. Fig. \ref{fig_survey_roadmap} shows the roadmap of this survey. As can be seen from the figure, we review and compare technologies including data streams, data storage models, searching, and event processing technologies, which play a vital role in enabling the vision of IoT. We also describe some relevant applications from several representative areas.
Although some surveys about IoT have been conducted recently (e.g., \cite{AtzoriIM10,ZengGC11,Perera-corr2013,Li-isf2014}), they focus on high level general issues and are mostly fragmented. In addition, these surveys do not specifically cover techniques on data processing and management, which is fundamentally critical to fully embrace IoT. To the best of our knowledge, this is the first survey that studies and discusses state-of-the-art techniques of IoT from the data-centric perspective.    

The remainder of the article is organized as follows. Section \ref{sec:data_streams} surveys the data streaming techniques and Section \ref{sec:data_models} focuses on the data models and storage technologies for IoT. Search and event processing technologies are discussed in Sections \ref{sec:searching} and \ref{sec:event_processing}, respectively. In Section \ref{sec:app}, some typical ongoing and/or visionary IoT applications where data techniques for IoT can bring significant changes are described. Finally, Section \ref{sec:openIssues} highlights some research open issues on IoT from the data perspective and Section \ref{sec:summary} offers some concluding remarks.

\section{Data Streams}
\label{sec:data_streams}

A data stream is a sequence of data objects, of which the number is potentially {\em unbounded}. A data stream may be continuously generated at a rapid rate. In the data stream, each data object can be described by a multi−dimensional attribute vector within a continuous, categorical, or mixed attribute space \cite{csur/SilvaFBHCG13}. There are some typical characteristics of data streams:

\begin{itemize}
\item Continuous arrival of data objects
\item Disordered arrival of data objects
\item Potentially unbounded size of a stream
\item Normally no persistence of data objects after being processed
\item Changing probability distributions of the unknown data generation process
\end{itemize}

Due to the excessive amount of data produced by all kinds of things in the era of IoT, data streams play an important role in data processing and analysis. This section will focus on related data stream research efforts that can help handle IoT data. Our discussions include general data stream processing, RFID data stream processing, and RDF triple stream processing.

\subsection{General Data Stream Processing}

Data streams can be generated in various scenarios, including a network of sensor nodes, a stock market or a network monitoring system and so on. In many scenarios such as the sensor network scenario, sensor nodes are normally powered by batteries or solar panels. Therefore, in typical a sensor data processing system, one of the challenging issues is power constraints. In most applications, communication across sensor networks or with a centralized server requires the largest amount of energy as sensing consumes less energy \cite{ads/SubramaniamG07}. If sensor nodes send their raw sensing data to a server without consideration of the amount of energy needed to communicate, the battery life of the sensor nodes could be drastically reduced. Consequently, sensor data processing techniques, including data aggregation, data compression, modeling and online querying, should be performed {\em on-site} or {\em in-network} to reduce communication cost \cite{ads/SubramaniamG07}. Furthermore, numerous demands on efficient data processing algorithms for sensor systems arise due to the limitations of computational power of sensor nodes as well as the existence of inaccuracy and bias in the sensor readings. In other scenarios, such as stock market and network monitoring systems, there also exist challenges in processing high-rate data streams.

\subsubsection{Query Processing} 
There are several important queries to be considered \cite{ads/SubramaniamG07}:

\begin{itemize}
\item {\em Aggregate Queries}. Aggregate Queries is an important class of queries in sensor systems, including \texttt{MIN}, \texttt{COUNT} and \texttt{AVG} operators. Various techniques have been proposed to efficiently process these aggregate operators in sensor systems, which can help to effectively reduce power consumption. Considering the properties of the aggregate functions, the in-network partial data could be preprocessed first, which can then be utilized to produce the final results for the issued queries.

\item {\em Join Queries}. An example of join queries is \textit{``Return the objects that were detected in both regions R1 and R2"} \cite{ads/SubramaniamG07}.  To evaluate the query, stream readings from the sensors in regions R1 and R2 should be joined first before we can determine whether an object was detected in the two designated regions. Join queries are useful in many applications, such as monitoring an environment where multiple sensors are deployed, tracking moving objects that are monitored by several types of sensors, etc.

\item {\em Top-k Monitoring}. \citeN{sigmod/BabcockO03} investigated the general problem of monitoring top-k values from distributed data streams. A technique is proposed to ensure the validity of the most recently communicated top-k answers by maintaining some specified arithmetic constraints at the stream sources. User specified error tolerance is also considered in order to provide high-quality answers. This technique can help reduce the overall communication cost between different sources. 

\item {\em Continuous Queries}. To monitor designated changes in an environment, sensors are typically required to answer queries in continuous manner. For instance, motion or sound sensors might be used to evaluate some continuous queries, such as ``Turn lights off if no motion is detected in area A in the past 10 minutes". When the query constraints are satisfied, the action of turning lights off could be automatically triggered by these sensors. If there are more than one continuous query evaluated over the same sensor readings, the storage and computation can be optimized by exploiting the fact that the sources of the queries and their partial results could overlap \cite{ads/SubramaniamG07}.

\end{itemize}

\subsubsection{Stream Mining} Stream mining can extract useful rules/information from data streams. The following lists some typical tasks for stream mining:

\begin{itemize}
\item {\em Clustering}. Clustering is the task of grouping a set of objects in such a way that objects in the same group (called a cluster) are more similar to each other than to those in other groups (clusters). Clustering techniques for data streams typically continuously cluster objects on memory constrained devices with some time limitations. Due to these restrictions, there are some requirements to consider when designing algorithms for clustering data streams \cite{books/daglib/0030859}: 
(i) providing clustering results via fast and incremental processing of data objects; 
(ii) rapidly detecting new clusters or changes of existing clusters; 
(iii) scaling to the potentially unbounded number of objects in data streams; 
(iv) providing a model representation that is consistently compact regardless the number of data objects;
(v) rapidly detecting the presence of outliers and acting accordingly; and 
(vi) dealing with different data types, such as XML trees, DNA sequences, GPS temporal and spatial information.

\item \textit{Classification}. Classification \cite{comsur/WangL11} uses prior knowledge to guide the partitioning process to construct a set of classifiers to represent the possible distribution of patterns. Basically, compared with clustering, classification is a supervised learning process whereas clustering is an unsupervised learning process. More formally, a typical classification algorithm can be defined as follows \cite{comsur/WangL11}: given a predefined classifier and two sets of data, labeled data and unlabeled data, the labeled data is used to train the classifier and the unlabeled data can then be classified by the trained classifier.

\item \textit{Outlier and Anomaly Detection}. In outlier and anomaly detection, the main task is to find data points that are most different from the remaining points in a given data set. Most existing outlier detection algorithms are based on the distance between every pair of points. The points that are most distant from all other points will be marked as outliers \cite{vldb/KnorrN98}. This kind of algorithms suffer from the same performance issue as they all run in $O(n^2)$ time. Hence, it is difficult to extend such approaches to distributed streaming data sets because points in those data sets normally arrive at multiple distributed end-points and must be processed incrementally.

\item \textit{Frequent Itemset Mining}. Frequent itemset mining is to find sets of items or values that co-occur frequently, or in other words, to find co-occurrence relationships in a transactional data set. Here a transactional data set refers to a data set where a set of items appear together in some specified context. Given a predefined support $s$, the goal in frequent itemset mining is to find all subsets of items that occur at least $s$ number of times, or in other words, that appear in at least $s$ transactional data sets at hand. Frequent itemset mining is both CPU and I/O intensive. Therefore, it is costly to completely re-mine a dynamic data set, which will be a typical case in IoT.

\end{itemize}

\subsubsection{Data Stream Processing in IoT} 

In IoT, multiple data streams processing \cite{cikm/GuoZTG11,cikm/XieZSzP12} would be more preferable as data streams can be generated at anywhere around the world and can be accessed globally via the Internet if being made public. For example, SmartSantander\footnote{http://www.smartsantander.eu/} proposes a city-scale experimental research facility in support of typical applications and services for a smart city. Around 20,000 sensors have been deployed to provide a variety of services, such as static environmental monitoring, mobile environmental monitoring, parks and gardens irrigation, outdoor parking area management, guidance to free parking lots and traffic intensity monitoring. A large number of data streams have to be processed efficiently to provide real-time monitoring of a smart city.

\subsection{RFID Data Stream Processing}
In 2003, a nonprofit open forum called the Ubiquitous ID Center\footnote{uID Center: www.uidcenter.org} was established. So far, more than 500 companies and organizations worldwide have contributed to it, publishing uID standards and industrial open standard specifications. uID standards are based on the uID architecture \cite{KoshizukaS10}, which identifies real-world entities via Radio-Frequency Identification (RFID) tags or barcodes, determines contextual information such as environmental parameters from networked sensors, and adapts information services according to the data it obtains.

RFID systems consist of radio frequency (RF) tags (also called transponders) and RF tag readers (also called transceivers). Readers may be able to both read data from and write data to a transponder. RFID is a promising electronic identification technology that enables real-time monitoring and tracking applications in a variety of domains. Streams of RFID data, whose basic form is a triplet  $<tag id$; $reader id$; $timestamp>$, raise new challenges since the data may be insufficient, incomplete, and voluminous~\cite{Sheng-Computer2008}.

\subsubsection{RFID Data Cleaning (Uncertainty and Unreliability)}

SMURF (Statistical sMoothing for Unreliable RFid data) \cite{vldb/JefferyGF06} is the first declarative, adaptive smoothing filter for cleaning raw RFID data streams. Unlike conventional techniques which expose the smoothing window parameter to the application, SMURF adapts the window size automatically and continuously over the lifetime of the system based on observed readings.

Periods of dropped readings and periods when a tag has moved are difficult to distinguish, which poses some challenges for the design of SMURF. To overcome such difficulty, a statistical sampling-based approach is put forward in SMURF. The main motivation is that RFID data streams can be modeled as a random sample of the tags in a reader's detection range. This sample-based view of observed RFID readings enables SMURF to develop algorithms based on statistical sampling theory to adapt the window size effectively. Basically, the false reads in RFID streams can be classified into two categories \cite{cikm/LiaoLCW11}:
\begin{itemize}

\item {\em Missing-Reads}. Though an RFID tag indeed locates in the range of a reader, it might not be read at all, thereby leading to a false prediction that the tag is not present. This may be caused by the weakness of RF signal, shortage of power, shield of signal between the tag and the reader, and the collision between tags. This type of errors is also referred as false negatives.

\item {\em Cross-Reads}. When an RFID tag locates outside the range of a reader, but it might be captured by this reader which leads another false prediction that the object is present in the scope of this reader (sometimes called {\em ghost reading}). Cross-reads may be arisen by the reflection of metal items, the abrupt strengthen of RF, and the change of antenna directions. This type of errors is also called false positives. 
\end{itemize}

SMURF cannot eliminate the cross-reads generated by physical factors. A kernel density-based probability cleaning method, called KLEAP, can be used to filter the cross-reads in RFID data streams \cite{cikm/LiaoLCW11}. KLEAP considers cross-reads as outliers, thus, the determination of cross-reads is transformed into the issue of detecting outliers on data streams. The density-based methods often perform well than the distance-based one, so KLEAP applies the density-based methods to detect cross-reads. It detects the exact positions of tags over the RFID data streams through examining the kernel densities of each tag captured by multiple readers.

\citeN{edbt/FazzingaFFP14} exploit the knowledge on the map of the real world and on the motility characteristics (such as the maximum speed) of the monitored objects, even if the users who analyze the data are typically acquainted with these aspects. From this knowledge of the domain, constraints can be naturally derived on the connectivity between pairs of locations (direct unreachability constraints) and/or on the time needed for reaching a location starting from another one (traveling-time constraints). These constraints can be used to discard interpretations of the data corresponding to inconsistent trajectories. Then a graph is built in the following way: its nodes correspond to pairs $<location, timestamp>$ and inside the graph, paths from source to target nodes one-to-one correspond to the valid trajectories in real word. Each node or edge is assigned a probability obtained by revising the a priori probability of the corresponding pair $<location, timestamp>$, so that the overall probability of a source-to-target path is the conditioned probability of the corresponding trajectory. In this way, trajectories of RFID-monitored objects can be cleaned.

\subsubsection{RFID Data Inference and Compression}

RFID data inference techniques are closely related to RFID data cleaning techniques because inference techniques will need to clean RFID data first and then they can infer to the high level information about the tagged objects, i.e., location and containment relationships. Since raw RFID data contains a large amount of redundancies, RFID data compression is also applied to reduce space requirements after inference results have been obtained. RFID data compression is a further step beyond inference, where compression is performed based on the results of inference to remove the redundant data.

\citeN{icde/TranSCNDS09} consider noisy, raw data streams from mobile RFID readers and employ a probabilistic approach to translate these streams into clean, rich event streams with location information. Their probabilistic model is built based on the mobility of the reader, object dynamics, and noisy readings. {\em Particle filtering} is used to infer clean information about object locations from raw streams captured from mobile RFID readers. 

The aforementioned data cleaning and inference techniques focus on smoothing over time, where containment relationships are not considered. Containment refers to {\em inter-object} relationships, e.g., containment between objects, cases, and pallets. Containment queries can be useful for enforcing packaging and shipping regulations. \citeN{pvldb/CaoSDS11} provide some examples of containment queries, such as ``raise an alert if a flammable item is not packed in a fireproof case" or ``verify that the food containing peanuts is never exposed to other food cases for more than an hour". They also observe that some known containment relations can be used to determine object locations by smoothing over these facts. For example, suppose that we can infer  that a specific set of objects have been packed in the same container. According to such knowledge, if one object in the container is read, all of the other objects must be in the same place. However, the fact is that the containment relationships are not known in advance. Therefore, a graphic model is proposed to infer containment relationships and to detect changes in containment relationships \cite{pvldb/CaoSDS11,tkde/NieCCDS12}.

\subsection{RDF Triple Stream Processing}
Linked Data is a method for publishing structured data and interlink such data to make it more useful\footnote{en.wikipedia.org}. It builds upon standard Web technologies such as HTTP, RDF and URIs and extends these technologies to share information. Linked Data can be understandable by computers. Data from different sources can be connected and queried in the form of Linked Data. Basically, Linked Data refers to a set of best practices to be followed in order to publish and link data on the Web, using the following basic principles\footnote{http://www.w3.org/DesignIssues/LinkedData.html}:

\begin{itemize}
\item Use URIs as names for things.
\item Use HTTP URIs so that people can look up those names.
\item When someone looks up a URI, provide useful information, using appropriate standards (RDF, SPARQL).
\item Include links to other URIs, so that more things can be discovered.
\end{itemize}

The concept of {\em Linked Stream Data} applies the Linked Data principles to streaming data, so that data streams can be published as part of the Web of Linked Data. Stream reasoning can provide the abstractions, foundations, methods and tools required to integrate data streams, the Semantic Web and reasoning systems. Substantial research efforts have been put forward, focusing on how to apply reasoning on streaming data, how to publish raw streaming data and connect them to the existing data sets on the Semantic Web, and how to extend the SPARQL query language to process streaming data \cite{semweb/ZhangDCC12}. These research efforts lay some foundations of semantic IoT technologies, facilitating machine-to-machine communication in IoT.

\subsubsection{Linked Stream Processing and Reasoning}
\citeN{semweb/SequedaC09} initiate efforts to apply the linked data principles to stream (sensor) data, so that this wealth of information could be easily included in the Linked Data cloud\footnote{http://linkeddata.org/}.

There are three typical streaming RDF/SPARQLS engines, including Streaming SPARQL \cite{esws/BollesGJ08}, SPARQLStream \cite{semweb/CalbimonteCG10}, C-SPARQL \cite{edbt/BarbieriBCG10}, and EP-SPARQL \cite{www/AnicicFRS11}. Each of these systems also proposes its own SPARQL extension for streaming data processing. In these studies, SPARQL has been extended to have sliding window operators for RDF stream processing. 

For example, Streaming SPARQL extends SPARQL to support window operators. But it does not consider performance issues, specially when designing the data structures. Further, it does not consider the sharing of computing states for continuous execution. Another example is SPARQLStream, which aims at enabling ontology-based access to streaming data. It defines a SPARQLStream language, which can be translated into another relational stream language based on mapping rules. 

C-SPARQL (Continuous SPARQL) \cite{edbt/BarbieriBCG10} attempts to facilitate reasoning upon rapidly changing information. In C-SPARQL, continuous queries are divided into static and dynamic parts and streaming data is transformed into non-streaming data within a specified window in order to apply standard algebraic operations, such as aggregate functions like \texttt{COUNT, COUNT DISTINCT, MAX, MIN} and \texttt{AVG}. The static parts will be loaded into relations, and the continuous queries are executed by processing the stream data against these relations. Event Processing SPARQL (EP-SPARQL), a language to describe event processing and stream reasoning, can be translated to ETALIS \cite{www/AnicicFRS11}, a Prolog-based complex event processing framework. First, RDF-based data elements are transformed into logic facts, and then EP-SPARQL queries are translated into Prolog rules. 
 
Different from the above approaches, CQELS \cite{semweb/PhuocDPH11} is a native streaming RDF/SPARQL system built from scratch. CQELS defines and implements a native processing model in the query engine. Its query execution framework can also dynamically adapt the query processor to changes in the input data. By using data encoding and caching of intermediate query results, CQELS reduces external disk access on large Linked Data collections. Some indexing techniques are also adopted to enable faster data access. Table \ref{tab:LinkedStreamProcessingAndReasoning_comparision} compares all these systems from various aspects.

\begin{acmtable}{358pt}
\centering
\(\begin{tabular}{|p{100pt}|p{30pt}|p{90pt}|p{40pt}|p{36pt}|}
  Approach & Native & Aggregation Support  & Reasoning Support  & SPARQL 1.1 Support \\
\hline
\hline
Streaming SPARQL \cite{esws/BollesGJ08} & No & Limited & Limited  & Limited   \\
\hline
SPARQLStream \cite{semweb/CalbimonteCG10} & No & Limited  & Limited  & Limited \\
\hline
C-SPARQL \cite{edbt/BarbieriBCG10} & No & Rich & Limited & Limited  \\
\hline
EP-SPARQL \cite{www/AnicicFRS11} & No & Limited & Rich & Limited  \\
\hline
CQELS \cite{semweb/PhuocDPH11} & Yes & Limited & Limited & Limited 
\end{tabular}\)
\caption{Comparisons of Linked Stream Processing and Reasoning.}
\label{tab:LinkedStreamProcessingAndReasoning_comparision}
\end{acmtable}

\subsubsection{Extracting RDF Triples from Unstructured Data Streams}
\citeN{semweb/GerberHBSUN13} point out that although the current Linked Open Data (LOD) cloud has tremendously grown over the last few years, it delivers mostly encyclopedic information (such as albums, places, kings, etc.) and fails to provide up-to-date information. Based on such observation, they develop RdfLiveNews, an approach that allows extracting RDF from unstructured (i.e., textual) data streams in a fashion similar to the live versions of the DBpedia\footnote{http://live.dbpedia.org/sparql} and LinkedGeoData\footnote{http://live.linkedgeodata.org/sparql} datasets. RdfLiveNews takes unstructured data streams as its input. It firstly removes duplicates in the streams. Then it uses the cleaned streams as a basis to extract patterns for relations between known resources. Next, the patterns will be clustered to labeled relations and finally will be used as a basis for generating RDF triples.

\section{Data Storage Models}\label{sec:data_models}
The nature of data produced by the Internet of Things calls for a revisit of data storage techniques, which will be further discussed in this section.

\subsection{New Architecture}
Traditional Database Management Systems (DBMSs) employ record-oriented (i.e., a record is represented by a row in a relational table) storage systems. With this row store architecture, a single disk write is able to store a single record with multiple attributes to disk. Records writes and updates are normally of high performance in these systems. Therefore, a DBMS with a row store architecture can be called a {\em write-optimized} system. In contrast, some systems need to deal with ad-hoc querying of large amounts of data, where read performance is of more importance. For such systems, {\em read-optimized} is the major design factor. Take data warehouses as an example. They represent one class of read-optimized system. In these read-optimized systems, a column-store architecture is a better choice. This is because in a column-store system, the values for each single column (or attribute) are stored contiguously, which can be easily optimized for high-performance querying.

\citeN{StonebrakerABCCFLLMOORTZ05} designed C-Store, a column-store architecture that supports the standard relational logical data model. Compared with the traditional DMBS architecture, the major differences are: (i) data in C-Store is not physically stored using its related relational logical data model; and (ii) whereas most row stores implement physical tables directly and then add various indexes to speed access, C-Store implements only projections. Here, projections are sorted subsets of the attributes of a table. Furthermore, \citeN{StonebrakerMAHHH07} show superior performance of column store based systems over the major RDBMS (relational DBMS) system. It is experimentally demonstrated that specialized engines in the data warehouse, stream processing, text, and scientific database markets can speed up the querying performance by 1-2 orders of magnitude using the column-store architecture. They also suggest that the DBMS vendors (and the research community) should start from scratch and design novel systems for requirements to be fulfilled in the near future, rather than just adapting current systems for those new requirements.

\subsection{Large-Scale Storage in Distributed Environments}

Storage issues in large scale systems have arisen due to the arrival of the big data era. For example, users of websites such as Facebook, Ebay and Yahoo! usually demand fast response times. One solution for this is to replicate data across globally distributed datacenters. However, \citeN{KadambiCCLRSTG11} discover that to replicate all data to all locations may waste huge amounts of resources since users from different locations may have different data consumption needs. For example, an European server may not need to maintain a replica of some rare accessed records in an Asian server. By exploiting such observations and selectively replicating large-scale web databases on a record-by-record basis, bandwidth and disk costs can be saved.

To meet the exceptional demands of data storage in IoT, developments of large-scale, distributed storage systems are of essential. There are three factors or requirements to be considered when designing a distributed storage system \cite{monet/ChenML14}:

\begin{itemize}
\item  {\it Consistency:} Consistency means to ensure that multiple copies of the same data are identical since server failures and parallel storage may cause inconsistency.

\item  {\it Availability:} Availability refers to the requirement that the entire distributed storage system (which contains multiple servers) should not be seriously affected by some extent of server failures and should be able to provide satisfactory reading and writing performance.

\item {\it Partition Tolerance:} Since multiple servers are interconnected by a network and the data is partitioned across the network, the distributed storage system should have a certain level of tolerance to problems caused by network failures. This refers to partition tolerance requirement.
\end{itemize}

Interestingly, it has been proved by \citeN{sigact/GilbertL02} that a distributed storage system could not simultaneously meet the requirements on consistency, availability, and partition tolerance, and at most {\em two of the three requirements} can be satisfied at the same time. On top of this theory, there are three types of distributed storage systems: (1) a CA system, which ignores partition tolerance; (2) a CP system, which ignores availability; and (3) an AP system, which ignores consistency. The comparisons of these systems and some of their representative works are summarized in Table \ref{tab:CAP_comparision}.

\begin{acmtable}{358pt}
\centering
\(\begin{tabular}{|p{30pt}|p{100pt}|p{100pt}|p{80pt}|}
  Type & Pros & Cons  & Representatives \\
\hline
\hline
CA 
& \begin{itemize}
\item Single copy of data
\item Consistency is easily ensured
\item Availability is assured by the excellent design of databases
\end{itemize}       
& \begin{itemize}
\item Could not handle network failures
\end{itemize} 
& \begin{itemize}
\item Traditional small-scale relational databases
\end{itemize} \\
\hline
CP 
& \begin{itemize}
\item Maintain several copies of the same data
\item A certain level of fault tolerance is ensured
\item Consistency is ensured by guaranteeing multiple copies of data to be identical
\end{itemize}       
& \begin{itemize}
\item Could not ensure sound availability due to the high cost for consistency assurance
\end{itemize} 
& \begin{itemize}
\item BigTable \cite{tocs/ChangDGHWBCFG08}
\item Hbase \cite{HBase}
\end{itemize} \\
\hline
AP 
& \begin{itemize}
\item Maintain several copies of the same data
\item A certain level of fault tolerance is ensured
\item Availability is assured by the design of distributed storage systems
\end{itemize}       
& \begin{itemize}
\item Strong consistency is not ensured
\item May cause a certain amount of data errors
\end{itemize} 
& \begin{itemize}
\item Dynamo \cite{sosp/DeCandiaHJKLPSVV07}
\item Cassandra \cite{sigops/LakshmanM10}
\end{itemize} 
\end{tabular}\)
\caption{Comparisons of three types of distributed storage systems.}
\label{tab:CAP_comparision}
\end{acmtable}

\subsection{Storage on Resource-Constrained Devices}
Storage issues also arise in resource-constrained scenarios in IoT. For example, in sensor networks, communication activity normally plays a more important role than storage. But \citeN{Mottola10} argues that for batch data collection, delay-tolerant mobile applications, and disconnected operations in static networks, the storage-centric paradigm becomes more critical. It is favored by decreasing costs and increasing capacity of storage hardware. SQUIRREL is also proposed in the same work, which is a lightweight run-time layer allocating data to different storage areas, based on data size versus energy trade-offs.

\citeN{YangWNLA09} developed SolarStore, a power storage service for solar-powered storage-centric sensor networks. The main goal of SolarStore is to improve the total amount of data that can be eventually retrieved from the network. It adaptively balances data reliability against data sensing since solar energy is renewable and dynamic. For example, it chooses to replicate data in the network until the next opportunity to upload data to the server. The degree of data replication also varies dynamically depending on the availability of solar energy and sensor storage.

Early database systems for sensor networks such as TinyDB and Cougar only act as filters for data collection networks and not as databases, i.e., no data is stored in, or retrieved from, any database. \citeN{TsiftesD11} presented a database management system for resource-constrained sensors named Antelope. Antelope supports run-time creation and deletion of databases and indexes and hence is a dynamic database system. It is the first DBMS for resource-constrained sensor devices, which enables a class of sensor network systems where every sensor holds a database. \citeN{TsiftesD11} also envisioned that database techniques would become increasingly important in the progress of sensor network applications and energy-efficient storage. Further, indexing and querying would play important roles in emerging storage-centric applications. 

Besides, \citeN{Nath09} proposes to use flash storage for logging data on a sensor node, called {\em amnesic storage} systems. An amnesic storage system archives streaming data using two key techniques: (i) data is compressed (usually with lossy compression methods) in an online fashion before being archived; and (ii) an amnesic storage system uses aging archived data by reducing the fidelity of older data to make space for newer data.

\section{Search Techniques}\label{sec:searching}

Searching and finding relevant objects from billions of things is one of the major challenges for the future Internet of Things and can bring about huge potential impact to humans~\cite{barnagi-is2013,ChristopheVT11,WangTL10,OstermaierRMFK10}. Supporting technologies for searching things in the IoT are very different from those used in searching Web documents because things are tightly bound to contextual information (e.g., location) and have no easily indexable properties (e.g., human readable text in the case of Web documents). In addition, the state information of things is dynamic and rapidly changing. Things discovery calls for innovative ways of managing and searching from dynamic data, which makes it different from traditional Web searching. This section overviews the relevant areas such as the Deep Web, Semantic Web and then discusses state-of-the-art techniques in searching things in the IoT environments.

\subsection{Deep Web and Semantic Web}
Deep Web refers to data stored outside Web pages and accessible from the Web, typically through HTML forms. The size of the Deep Web is estimated to be several orders of magnitude larger than that of the so-called {\em Surface Web} (the Web that is accessible and indexable by text search engines) \cite{CaliM10}.

Executing structured, high-level queries on deep web data sources involves a number of challenges because query execution engines have a very limited access to data \cite{WangA11}. Besides, hidden data on the Deep Web may not provide the domain information for an attribute. Hence, \citeN{JinZD11} argue that {\em domain discovery} becomes a critical challenge as a broad range of existing techniques on third-party analytical and mash-up applications\footnote{Here, a mash-up is a Web page or application that integrates complementary elements from two or more sources.} are being applied over hidden databases. Furthermore, the traditional way to access hidden data on the Deep Web (by manually filling-up HTML forms on search interfaces) is not scalable given the growing size of the Deep Web. Therefore, \citeN{KhareAS10} argue that automatic access to such hidden data requires an automatic understanding of search interfaces by computers, which would be challenging, as the interfaces are originally designed for human access.

The Deep Web provides a wealth of hidden data in semi-structured form, accessible through Web forms and Web services. Since the data is hidden, to reach the whole content of the World Wide Web by just following hyperlinks is impossible. Regarding such issues, on top of XML, the Semantic Web grows as a common structured data source \cite{SuchanekVNS11}. With the W3C standards Resource Description Framework (RDF) and Web Ontology Language (OWL), the Semantic Web aims to unify the way semantic information is stored and exchanged. The Semantic Web makes it possible for machines themselves to not just read, but also ``understand" the data from data sources, which enables machine to machine communication. In particular, languages such as as Microformats\footnote{http://www.microformats.org} and schema.org, can be used to add semantics to the descriptions of Web resources (including things).

\subsection{Web Search}
The frequent changes and the unprecedented scale of the Web together pose enormous challenges to Web search engines, making it challenging to provide the most up-to-date and highly relevant information to its users \cite{ChoG10}. In IoT, this may become even more challenging as things would scale up the Web further and make the Web change more rapidly. For example, Tsubuyaku Sensor\footnote{http://ts.uctec.com/tsensor/index-e.php} is a new wireless device from Japanese Ubiquitous Computing Technology. It can monitor conditions such as temperature, humidity and radiation levels. It then automatically tweet the resulting data via Twitter. In this way, a sensor becomes a virtual Twitter user, which can actively post tweets on the Web. 

\subsubsection{Information Extraction}
Information extraction from the Web is of growing importance. For example, objects on the Web are often associated with many attributes that describe the objects, making it essential to extract such attributes and map them to their corresponding objects. However, much attribute information about an object is hidden in the dynamic user interaction and is not on the Web page that describes the object. \citeN{HuangWJF11} build a search model for exploratory Web sites, and algorithms are also proposed for identifying, clustering, and relationship mining of related Web pages based on the model. Besides, \citeN{SatpalBSRS11} study the problem of extracting structured data from web pages taking into account both the content of individual attributes as well as the structure of pages and sites. Markov Logic Networks (MLNs) are adopted to capture both content and structural features in a single unified framework, which is able to introduce more accurate inference. 

The corpus of a search engine forms a rich source of information of analytical interest to third parties, but the only available access is by issuing search queries through its interface. \citeN{ZhangZD11} claim that, in order to support data analytics over a search engine's corpus, it is necessary to address two main problems: (i) the sampling of documents (for offline analytics); and (ii) the direct (online) estimation of aggregates. Meanwhile, in order to complete data analytics tasks, only a small number of queries would be issued to a search engine through the keyword-search interface.

\subsubsection{Real-time Web Search}
Real-time web search refers to the retrieval of very latest content which is in high demand. It is reported that Twitter handled more than 50 million tweets per day. Providing real-time search service is indeed very challenging in such large-scale microblogging systems because thousands of new updates need to be processed per second \cite{ChenLOW11}.

\citeN{SakakiOM10} observe that Twitter is real-time micro-blogging and investigate the real-time interaction of events such as earthquakes in Twitter. They consider each Twitter user as a virtual sensor and apply Kalman filtering and particle filtering for estimating the centers of earthquakes and the trajectories of typhoons. Similarly, \citeN{DongZKBDCZZ10} identify two challenges not encountered in non-real-time web search when supporting real-time web search, which are (i) quickly crawling relevant content and (ii) ranking documents with link and click information. Then they propose to use the micro-blogging data stream to detect fresh URLs and to compute novel and effective features for ranking fresh URLs based on micro-blogging data.

\subsubsection{Searching information over RDF Data} Searching information from RDF data is important as more and more information is published in the form of RDF (e.g., via Linked Open Data Cloud). Efficient management of RDF data is also an important factor in realizing the Semantic Web vision \cite{AbadiMMH07}. Performance and scalability issues need to be addressed as the Semantic Web technology is applied to real-world applications. Unlike the relational database community, the Semantic Web community uses a very different data model, which is RDF.

\citeN{TsatsanifosSS11} present MIDAS-RDF, a distributed P2P RDF/S repository that is built on top of a distributed multi-dimensional index structure. It features fast retrieval of RDF triples satisfying various pattern queries by translating them into multi-dimensional range queries, which can be processed by the underlying index in hops logarithmic to the number of peers. Further, since in IoT, data uncertainty is critical, \citeN{LianC11} address the problem of efficiently answering queries on probabilistic RDF data graphs. They model RDF data by probabilistic graphs, and an RDF query is equivalent to a search over subgraphs of probabilistic graphs that have high probabilities to match with a given query graph.

\subsubsection{Collaborative Web Search}
Web search engines often answer user queries based on data and information in relevant structured databases, which will be searched in isolation. Since a single database may not contain sufficient information to answer the query, the search often produces empty or incomplete results. Motivated by this observation, \citeN{AgrawalCCGKX09} exploit web search results and the items in structured databases together to produce more complete answers to a wide range of queries that traditional web search cannot support well. Take query ``light-weight gaming laptop'' as an example. Dell XPS M1330 should be considered a match to such query as it is a light-weight laptop and suitable for gaming. But if searching only the query keywords \{\texttt{light-weight, gaming}\} on the Web, Dell XPS M1330 may not appear in the search results. Therefore, \citeN{AgrawalCCGKX09} propose to utilize the web search results (e.g. a set of relevant web documents) to help identify relevant information in some structured databases. Then the user queries could be better answered.

Similarly, \citeN{ChaudhuriGX09} exploit web search engines in order to define new similarity functions for recognizing named entities such as products, people names, or locations from documents, such as ``X61" and the entity ``Lenovo ThinkPad X61 Notebook". The proposed new similarity functions are more accurate than existing string-based similarity functions because they aggregate evidence from multiple documents, and exploit web search engines to measure similarity.

\subsection{Search of Things in IoT}
In IoT, connecting things enabled by RFID, embedded sensors and sensor networks to the Internet and publishing their output on the Web would become a reality. Real-world objects would have their own Web presence. Considering the potential and profound impact of IoT technologies, search of things in IoT will become as important as today's document search on the Web \cite{OstermaierRMFK10}.

\subsubsection{Key Words based Search of Things}

Unlike search engines such as Google, searching for information in the physical world is more difficult because the physical objects do not have (reliable) connections to virtual space. For example, online books can be easily discovered by searching but physical books at home may be more difficult to find. This observation motivates \citeN{WangTL10} to propose Snoogle, a search engine for the physical world. The basic idea behind Snoogle is that sensor nodes carry a textual description of the object they will be attached to. Such description forms the keywords for search of things. Then the key words information of the whole sensor network is indexed using a two-tiered hierarchy. The lower tier contains many mediators, which are also called {\em index points}. Each index point maintains an aggregate view of all sensors in a local area (e.g. a room) and every sensor in the same area will be assigned to the same index point. In the top tier, there is a single mediator called the {\em key index point}. The key index point will maintain an aggregate view of the whole network.

MAX, a system that users can easily locate objects, is also designed \cite{YapSM08}. The main assumption is that tags are attached to everyday objects and each tag stores a descriptor of the object it is attached to (e.g., the book of \texttt{Harry Potter}). Multiple descriptor words are allowed in each tag, enabling users to label the object with richer information, so that others can locate the object based on the label. A three-tiered hierarchy of mediators is used. In the lowest tier, substations represent immobile objects such as tables or shelves, on which mobile tagged objects can be placed. In the middle tier, base stations represent a geographical space such as a room containing multiple substations. In the top tier, the MAX server represents the entire space covered by the system. When searching for an object left behind, it is easy to locate where the object has been left by exploiting the knowledge of which substation and base station it belongs to. 

Microsearch is a system that runs on resource constrained small devices capable of being embedded into everyday objects \cite{TanSWL10}. It allows users to do textual search in the local storage of a stand-alone small device, without support from a backend server. The challenge is that Microsearch runs in a resource constrained platform, where conventional search engine design and algorithms cannot be used. \citeN{TanSWL10} also propose information retrieval (IR) techniques for query resolution which can answer top-k queries in a space-efficient manner.

Another search engine mainly designed for searching things, called Dyser, is proposed in \cite{ElahiROFK09,OstermaierRMFK10}. Dyser allows users to search for real-world entities with a given state, such as ``hot'' or ``cold''. However, this approach imposes two strong conditions: (i) to perform a query, end users have to know the vocabulary used by sensors (how states are named); and (ii) an entity must be represented by all the sensors that compose it. In order to estimate the probability of a sensor matching a query with sufficient accuracy and to rank sensor matching results, prediction models are adopted. The key idea of sensor ranking is to exploit the periodic nature of people-centric sensors by using appropriate prediction models.

\subsubsection{Collaborative Search of Things}
\citeN{FrankBMK08} present a comprehensive system for managing and finding everyday objects relying on the collaboration of mobile phones in an urban area as object-sensing devices. For such tasks, the authors argue that the necessary infrastructures for such system include a sensing infrastructure, a communication infrastructure and a commercial infrastructure. Because of these requirements, the modern mobile phone system, which contains mobile sensors, provides a unique opportunity to realize collaborative search of everyday objects. The sensing model of the proposed system associates a probability with locations, meaning that the object currently has a certain probability of being at a certain location, thereby accelerating the search speed and reducing communication cost. Mobility provided by mobile sensors increases spatial coverage and hence the probability of finding a sought object.

\section{Complex Event Processing}\label{sec:event_processing}

Data streaming techniques typically process incoming data through a sequence of transformations based on common SQL operators, like \texttt{selection}, \texttt{aggregate}, \texttt{join}, and these operators are defined in general by relational algebra. By contrast, the {\em complex event processing} (CEP) model views the information in the streams as events in the physical world. These events must be filtered, combined and transformed into higher-level events for better understanding by computers and humans \cite{csur/CugolaM12}. Similar to traditional publish-subscribe systems, CEP systems allow subscribers to express their interest in composite events. The focus of CEP model is on detecting occurrences of particular patterns of (low-level) events indicating some higher-level events, which may be interesting to some particular event subscribers. In the era of IoT, CEP techniques lay part of the foundation of supporting computers to sense and react to events in the physical world.

\subsection{Complex Event Processing}
Systems for event processing and in particular event recognition ({\em event pattern matching}) accept a stream of time-stamped, simple or low-level events as input. A low-level event is the result of applying a computational derivation process to some other event, such as an event coming from a sensor. Using low-level events as input, (complex) event processing systems identify composite or high-level events of interest \cite{edbt/ArtikisP14}. They are also collections of events that satisfy certain patterns.

SASE is a complex event processing system designed for monitoring queries over streams of RFID readings \cite{sigmod/WuDR06}. The SASE defines its own declarative event language that combines filtering, correlation, and transformation of events. The overall structure of the SASE language contains the \texttt{EVENT} clause specifying event patterns, the \texttt{WHERE} clause specifying qualifications and the \texttt{WITHIN} clause specifying window sizes. To meet the needs of RFID-enabled monitoring applications, several operators are also defined, including the \texttt{ANY} operator, the \texttt{SEQ\_} operator, the \texttt{SEQ\_WITHOUT} operator, the \texttt{Selection} operator and the \texttt{WITHIN\_} operator. In order to process SASE queries, a query plan in SASE adopts a subset of six operators: 
\texttt{sequence scan}, \texttt{sequence construction}, \texttt{selection}, \texttt{window}, \texttt{negation}, and \texttt{transformation}. Pipelined execution of the above operators is used. More specifically, if a query matches a current event and some previous events, these events will be emitted from sequence scan and sequence construction immediately and form an event sequence. This event sequence is then pipelined through the subsequence operators, and added to the final output. 
To realize sequence scan, the basis of the whole process, Non-deterministic Finite Automata (NFA) are used.

Pattern matching over streams has been studied by \citeN{sigmod/AgrawalDGI08}. It presents two new challenges: (i) compared with languages for regular expression matching, languages for pattern matching over streams are significantly richer; and (ii) the conventional techniques for stream query processing are inadequate for efficient evaluation of pattern queries over streams. In order to represent each pattern query, a new query evaluation model is designed for processing pattern matching over RFID streams, employing a new type of automaton that comprises a nondeterministic finite automaton (NFA) and a match buffer, named NFA$^b$ \cite{sigmod/AgrawalDGI08}. Because of the powerful expressiveness of NFA, the semantics for the complete set of event pattern queries can be captured by the NFA$^b$ model. Optimizations and query evaluation plans can also be produced and applied based on this model over event streams.

\begin{figure}[h]
\centering
\scalebox{1}{\includegraphics{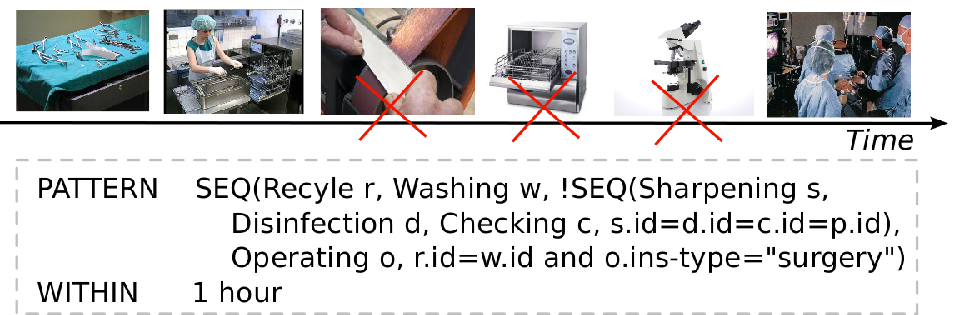}}
\caption{Nested CEP Query Example} (Adapted from \cite{icde/LiuRDGWAM11})
\label{fig_neel_example}
\end{figure}

\citeN{icde/LiuRDGWAM11} propose nested CEP language called NEEL to support the flexible nesting of \texttt{AND}, \texttt{OR}, \texttt{Negation} and \texttt{SEQ} operators at any level. One NEEL query example is given in Fig. \ref{fig_neel_example}, which expresses ``a critical condition that after being recycled and washed, a surgery tool is being put back into use without first being sharpened, disinfected and then checked for quality assurance" \cite{icde/LiuRDGWAM11}. Several techniques are also proposed to accelerate the evaluation of nested queries. Firstly, nested event expressions will be converted into normal forms by a  normalization procedure. Secondly, a group of similar sub-expressions will be processed using prefix caching, suffix clustering methods and a customized physical execution strategy. Thirdly, an optimizer for optimal shared execution method is also designed based on the idea of iterative improvement. Compared with the traditional iterative nested execution, the optimized NEEL execution is up to two orders of magnitude faster. \citeN{edbt/RayRLGWA13} optimize the processing of Nested Complex Event Processing queries by designing the Continuous Sliding View structures for inner sub-queries.

Recent efforts have also been put on other aspects of complex event processing. 
For example, \citeN{vldb/HeinzeJPGJF13} study complex event processing in a distributed environment and propose FUGU -- an elastic allocator for Complex Event Processing systems. FUGU can dynamically allocate and de-allocate both stateless and stateful queries in order to meet the utilization goals. To that end, FUGU relies on bin packing to allocate queries to hosts. Very recently, \citeN{icdt/HeBN14} investigate load shedding techniques for complex event processing under various resource constraints. Like other stream systems, CEP systems often face bursty input data. Since over-provisioning the system to the point where it can handle any such burst may be uneconomical or impossible, during peak loads a CEP system may need to ``shed" portions of the load. The key technical challenge is to selectively shed work in order to eliminate the less important query results, thereby preserving the more useful query results defined by some utility function. Motivated by this, several load shedding algorithms are designed, including CPU-bound load shedding, memory-bound load shedding, and dual-bound load shedding (with both CPU- and memory-bound), depending on which resource is constrained.

\subsection{Semantic Complex Event Processing}

The combination of event processing and knowledge representation can lead to novel semantic-rich event processing engines \cite{debs/TeymourianP09,debs/ZhouSP11,debs/TeymourianRP12}. These intelligent event processing engines can (i) help to understand what is happening in terms of events, (ii) state and know what reactions and processes it can invoke, and furthermore (iii) decide what new events it can signal. The identification of critical events and situations requires processing vast amounts of data and meta-data within and outside the systems.

\subsubsection{Semantic CEP System}

\begin{figure}[!h]
\centering
\scalebox{0.5}{\includegraphics{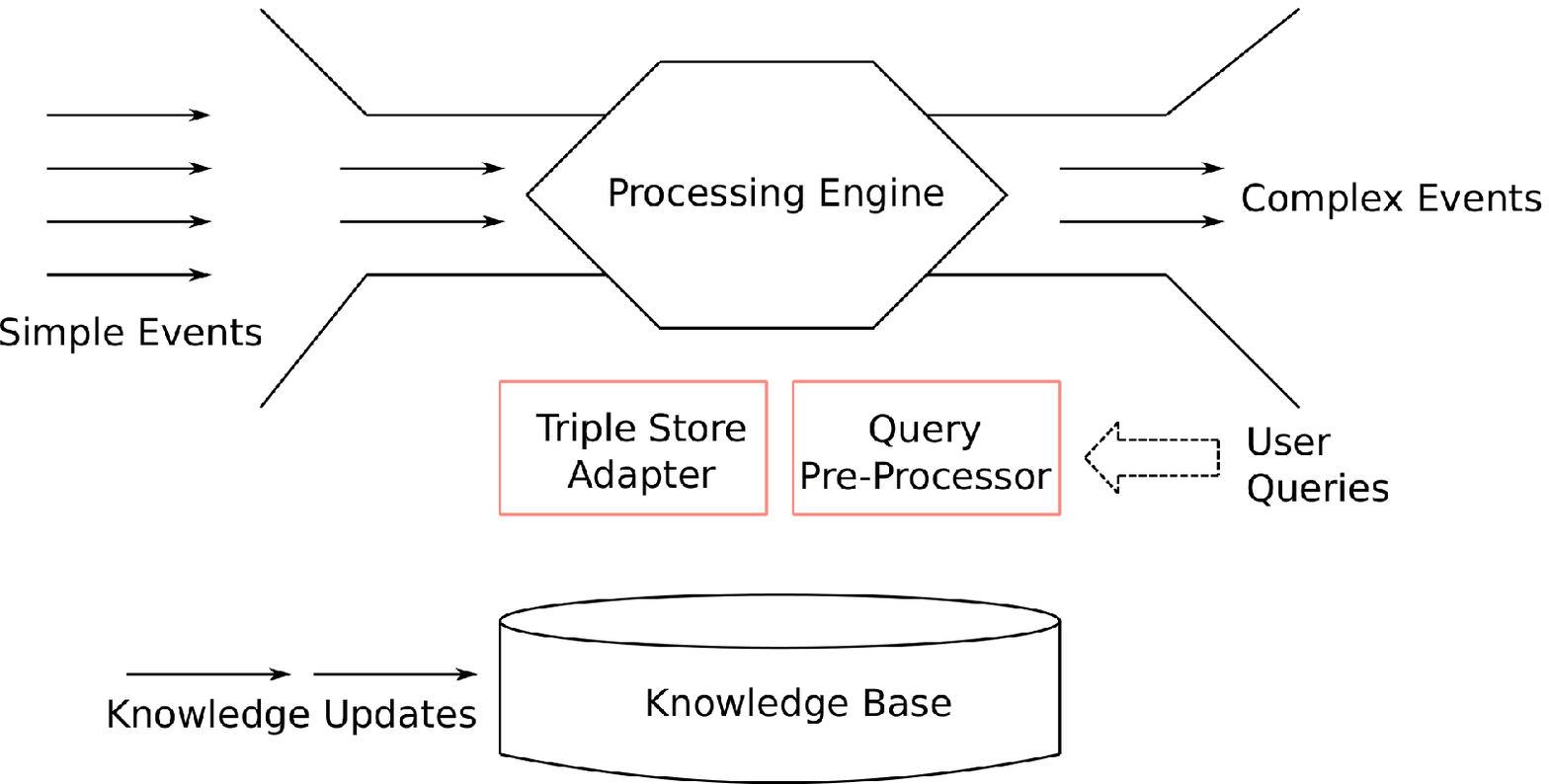}}
\caption{Semantic Complex Event Processing System Overview} (Adapted from \cite{debs/TeymourianRP12})
\label{fig_scep_sys}
\end{figure}

A semantic CEP system is shown in Fig. \ref{fig_scep_sys}. Semantic models of events can improve event processing quality by using event meta-data in combination with ontologies and rules (i.e., {\em knowledge bases}). The fusion of background knowledge with data from an event stream can help the event processing engine to know more about incoming events and their relationships to other related concepts. A Knowledge Base (KB) can be used to provide background knowledge about the events and other non-event resources \cite{debs/TeymourianRP12}. This means that events can be detected based on reasoning on their type hierarchy, temporal/spatial relationships, or their relationship to other objects in the application domain. 

The benefits of using background knowledge in complex event processing can be seen as two major advantages over state-of-the-art CEP systems. The first benefit is its higher expressiveness and the second one its flexibility. Expressiveness means that an event processing system can precisely express complex event patterns and reactions to events which can be directly translated into business operations. Flexibility means that a CEP system is able to integrate new business changes into the systems in a fraction of time rather than changing the whole event processing rules. Complex event patterns are independent of current businesses and are defined in a higher level of abstraction based on business strategies. When something is changed in the business environment, it can be considered simply as an update in the background knowledge and the complex event detection patterns which are defined based on the business plans should not be changed.

\subsubsection{Semantic Event Enrichment}
The usage of background knowledge about events and their relations to other concepts in the application domain can improve the expressiveness and flexibility of CEP systems. Huge amounts of domain background knowledge stored in external knowledge bases can be used in combination with event processing in order to achieve more knowledgeable complex event processing. \citeN{edbt/TeymourianRP12} discuss the benefits of background knowledge for event processing and describe different categories of event query rules. 

\citeN{debs/HasanOC13} identify an information completeness problem in semantic event processing contexts from a different angle. For example, while the basic information item in an event-based system is an event, normal users often require the system to handle information that is not encoded in the event. Such information typically comes from legacy databases or web data sources. This requires some degrees of information completeness or incompleteness for events to be sufficient for tasks such as subscription matching. The process of reducing information incompleteness is called {\em event enrichment}. Several challenges are identified for event enrichment, including determination of the enrichment source, retrieval of information items from the enrichment source, finding complementary information for an event in the enrichment source and fusion of complementary information with the event. To address these challenges, a model based on unifying enrichment within the event consumer logic and a native enricher that tackles incompleteness before matching are proposed \cite{debs/HasanOC13}.

\subsubsection{Approximate Semantic Matching}
Approximate semantic matching is first studied by \citeN{debs/ZhouSP11}. To achieve approximate matching, semantic selection and inexact selection are used. More specifically, the semantic selection evaluates pattern constraints based on the semantic equivalence of attribute meanings captured by the event ontology instead of syntactic identical attribute values, while the inexact selection selects events and allows a limited number of mismatches to detect relevant patterns. A similarity function is associated with the inexact selection to evaluate relevance between matching patterns and target patterns.

\citeN{debs/HasanOC12} study approximate semantic matching of heterogeneous events. The motivation is that heterogeneous events are difficult to match in a distributed computing environment as similar or closely related events may not be described using the same words but in a semantically related form. To match all interesting events, users may have to write many slightly different subscriptions and have to know the exact format of all the heterogeneous events. Based such observation, semantic decoupling of events and user's subscriptions becomes necessary. However, after such decoupling, the subscriptions would hardly exactly match the descriptions of events. This indicates that approximate matching and processing of events are inevitable.
A model for approximate semantic matching that addresses event semantic decoupling is proposed. The model is evaluated using a hybrid matching approach based on both thesauri, semantic similarity and relatedness measures. After adopting this technique, the number of event subscriptions to achieve sufficiently precise matching results can be greatly reduced because of the decoupling between events and user subscriptions.

\section{Potential IoT Applications}\label{sec:app}

As \citeN{AshtonK09} points out that IoT ``has the potential to change the world, just as the Internet did''. 
The ongoing and/or visionary IoT applications show that IoT can bring significant changes in many domains, i.e. cities and homes, environment monitoring, health, energy, and business, etc. \citeN{MatternF10} also argue that IoT can bring the ability to react to events in the physical world in an automatic, rapid and informed manner. This also opens up new opportunities for dealing with complex or critical situations and enables a wide variety of business processes to be optimized. In this section, we overview several representative domains where IoT can make some profound changes. 

\subsection{Smart Cities and Homes}

IoT can connect billions of smart things and can help capture information in cities. Based on IoT, cities would become smarter and more efficient. Below are some examples of promising IoT applications in future smart cities.
In a modern city, lots of digital data traces are generated there every second via cameras and sensors of all kinds  
\cite{Guinard10}. All this data represents a goldmine for everyone, if people in the city would be able take advantage of it in an efficient and effective way. For example,  IoT can facilitate resources management issues for modern cities. Specifically, static resources (e.g. fire stations, parking spots) and mobile resources (e.g. police cars, fire trucks) in a city can be managed effectively using IoT technologies \cite{GaoGMZ09}. Whenever events (fires, crime reports, cars looking for parking) arise, IoT technologies would be able to quickly match resources with events in an optimal way based on the information captured by smart things, thereby reducing cost and saving time. 
Taxi drivers in the city would also be able to better serve prospective passengers by learning passenger's mobility patterns and other taxi drivers' serving behaviors through the help of IoT technologies \cite{YuanZZXS11}. Besides, one study estimated a loss of \$78 billion in 2007 in the form of 4.2 billion lost hours and 2.9 billion gallons of wasted gasoline in the United States alone \cite{MathurJKCXGT10}. IoT could bring fundamental changes in urban street-parking management, which would greatly benefit the whole society by reducing traffic congestion and fuel consumption \cite{puc/Sujith14}.

Security in a city is of great concerns, which can benefit a lot from the development of IoT technologies. Losses resulted from property crimes were estimated to be \$17.2 billion in the U.S. in 2008 \cite{GuhaPLMKDK10}. Current security cameras, motion detectors, and alarm systems are not able to help track or recover stolen property. IoT technologies can help to deter, detect, and track personal property theft since things are interconnected and can interact with each other. IoT technologies can also help improve stolen property recovery rates, and disrupt stolen property distribution networks. Similarly, a network of static and mobile sensors can be used to detect threats on city streets and in open areas such as parks \cite{LiuBC11}.

With IoT technologies, people can browse and manage their homes via the Web. For example, they would be able to check whether the light in their bedrooms is on and could turn it off by simply clicking a button on a Web page. Similar operations and management could be done in office environments. Plumbing is ranked as one of the ten most frequently found problems in homes \cite{LaiCHC10}. It is important to determine the spatial topology of hidden water pipelines behind walls and underground. In IoT, smart things in homes would be able to report plumbing problems automatically and report to owners and/or plumbers for efficient maintenance and repair.

\subsection{Environment Monitoring}

IoT technologies can also help to monitor and protect environments thereby improving human's knowledge about environments. Take water as an example. \citeN{DetweilerDJSCR10} propose that understanding the dynamics of bodies of water and their impact on the global environment requires sensing information over the full volume of water. In such context, IoT technologies would be able to provide effective approaches to study water. Also IoT could improve water management in a city. Drinking water is becoming a scarce resource around the world. In big cities, efficiently distributing water is one of the major issues \cite{Guinard10}. Various reports show that on average 30\% of drinkable water is lost during transmission due to the aging infrastructure and pipe failures. Further, water can be contaminated biologically or chemically due to inefficient operation and management. In order to effectively manage and efficiently transport water, IoT technologies would be of great importance.

Soil contains vast ecosystems that play a key role in the Earth's water and nutrient cycles, but scientists cannot currently collect the high-resolution data required to fully understand them. Many soil sensors are inherently fragile and often produce invalid or uncalibrated data \cite{RamanathanSKWHE09}. IoT technologies would help to validate, calibrate, repair, or replace sensors, allowing to use available sensors without sacrificing data integrity and meanwhile minimizing the human resources required.

Sound is another example where IoT technologies can help. Sound is multidimensional, varying in intensity and spectra. So it is difficult to quantify, e.g., it is difficult to determine what kind of sound is noise. Further, the definitions and feelings of noise are quite subjective. For example, some noises could be pleasant, like flowing water, while others can be annoying, such as car alarms, screeching breaks and people arguing. \citeN{ZimmermanR11} design and build a device to monitor residential noise pollution to address the above problems. Firstly, noise samples from three representative houses are used, which span the spectrum of quiet to noisy neighborhoods. Secondly, a noise model is developed to characterize residential noise. Thirdly, noise events of an entire day (24 hours) are compressed into a one minute auditory summary. Data collection, transmission and storage requirements can be minimized in order to utilize low-cost and low-power components, while sufficient measurement accuracy is still maintained.

Intel has developed smart sensors that can warn people about running outside when the air is polluted\footnote{http://www.fastcoexist.com/1680111/intels-sensors-will-warn-you-about-running-outside-when-the-air-is-polluted}. For example, if someone is preparing to take a jog along his/her regular route, an application on his/her smartphone pushes out a message: air pollution levels are high in the park where he/she usually runs. Then he/she could try a recommended route that is cleaner. Currently, many cities already have pollution and weather sensors. They are usually located on top of buildings, far from daily human activities. 

\subsection{Health}

In future IoT environments, an RFID-enabled information infrastructure would be likely to revolutionize areas such as healthcare, and pharmaceutical \cite{GarfinkelR05}. For example, a healthcare environment such as a large hospital or aged care could tag all pieces of medical equipment (e.g., scalpels, thermometers) and drug products for inventory management. Each storage area or patient room would be equipped with RFID readers that could scan medical devices, drug products, and their associated cases. Such an RFID-based infrastructure could offer a hospital unprecedented near real-time ability to track and monitor objects and detect anomalies (e.g., misplaced objects) as they occur.

As personal health sensors become ubiquitous, they are expected to become interoperable. This means standardized sensors can wirelessly communicate their data to a device many people already carry today (e.g., mobile phones) \cite{CorneliusK11}. \citeN{LesterHPLBD09} argue that one challenge in weight control is the difficulty of tracking food calories consumed and calories expended by activity. Then they present a system for automatic monitoring of calories consumed using a single body-worn accelerometer. To be fully benefited from such data for a large body of people, applying IoT technologies in such area would be an promising direction.

Mobile technology and sensors are creating ways to inexpensively and continuously monitor people's health. Doctors may call their clients to schedule an appointment,rather than vice-versa, because the doctors could know their clients' health conditions in real-time. Some projects for such purpose have been initiated. For example, EveryHeartBeat\footnote{http://join.everyheartbeat.org/} is a project for Body Computing to ``connect the more than 5 billion mobile phones in the world to the health ecosystem". In the initial stage, heart rate monitoring is investigated. Consumers would be able to self track their pulse and studies show heart rate monitoring could be useful in detecting heart conditions and enabling early diagnosis. The future goal is to include data on blood sugar levels, and other biometrics collected via mobile devices. 

\subsection{Energy}

Home heating is a major factor in worldwide energy use \cite{ScottBKMHHV11}. In IoT, home energy management applications could be built upon embedded Web servers \cite{PriyanthaKGZ08}. Through such online web services, people can track and manage their home energy consumption. \citeN{GuptaIL09} present a system for augmenting these thermostats using just-in-time heating and cooling based on travel-to-home distance obtained from location-aware mobile phones. The system makes use of a GPS-enabled thermostat which could lead to savings of as much as 7\%. In IoT, as things in homes would become smart and connected to the Internet, similar energy savings could be more effective. For example, by automatically sensing occupancy and sleep patterns in a home, it would be possible to save energy by automatically turning off the home's HVAC (heating, ventilation, and air conditioning) system \cite{LuSSGHSFW10}.

Besides home heating, fuel consumption is also an important issue. \citeN{GantiPANA10} develop GreenGPS, a navigation service that uses participatory sensing data to map fuel consumption on city streets. GreenGPS would allow drivers to find the most fuel-efficient routes for their vehicles between arbitrary end-points. In IoT, fuel consumption would be further reduced by enabling cars and passengers to communicate with each other for ride sharing~\cite{YuanZZXS11}.

\subsection{Business}

IoT technologies would be able to help to improve efficiency in business and bring other impacts on business \cite{MatternF10}:

\begin{itemize}
  \item From a commercial point of view, IoT can help increase the efficiency of business processes and reduce costs in warehouse logistics and in service industries. This is because more complete and necessary information can be collected by interconnected things. owing to its huge and profound impact on the society, IoT research and applications can also trigger new business models involving smart things and associated services.
  \item From a social and political point of view, IoT technologies can provide a general increase in the quality of life for the following reasons. Firstly, consumers and citizens will be able to obtain more comprehensive information. Secondly, care for aged and/or disabled people can be improved with smarter assistance systems. Thirdly, safety can be increased. For example, road safety can be improved by receiving more complete and real-time traffic and road condition information.
  \item From a personal point of view, new services enabled by IoT technologies can make life more pleasant, entertaining, independent and also safer. For example, business taking advantages of technologies of search of things in IoT can help locate lost things quickly, such as personal belongs, pets or even other people.
\end{itemize}

Besides, take improving information handover efficiency in a global supply chain as an example. \citeN{StephanMKFO10} propose the {\em digital object memories} (DOM), which can store order-related data via smart labels on the item. Based on DOM, relevant life cycle information could be attached to the product itself. Considering the potential different stakeholders including manufacturer, distributor, retailer, and end customer along the supply/value chain, this approach facilitates information handover.

Further, there are many important bits of information in an IoT-based supply chain, such as the 5W (what, when, where, who, which). It is also necessary to integrate them efficiently and in real-time in other operations. The EPCIS (Electronic Product Code Information System) network is a set of tools and standards for tracking and sharing RFID-tagged products in IoT. However, much of this data remains in closed networks and is hard to integrate \cite{tpds/Wu13}. IoT technologies could be used to make it easier to use all this data, to integrate it into various applications, and to build more flexible, scalable, global application for better (even real-time) logistics.

\section{Open Issues}\label{sec:openIssues}

The development of IoT technologies and applications is merely starting off. Many new challenges and issues have not been addressed, which require substantial efforts from both academia and industry. In this section, we identify some key directions for future research and development from a data-centric perspective.



\begin{itemize}
  \item {\it Data Quality and Uncertainty:} In IoT, as data volume increases, inconsistency and redundancy within data would become paramount issues. \citeN{FanGMM10} indicate that one of the central problems for data quality is {\em inconsistency detection} and when data is distributed, the detection would be far more challenging. This is because inconsistency detection often requires shipping data from one site to another. Meanwhile, inherited from RFID data \cite{CaoSDS11} and sensor data \cite{PrabhakarC09}, IoT data would be of great uncertainty, which also presents significant challenges.

  \item {\it Co-Space Data:} In an IoT environment, the physical space and the virtual (data) space co-exist, and interact simultaneously. Novel technologies must be developed to allow data to be processed and manipulated seamlessly between the real and digital spaces \cite{OoiTT09}. To synchronize data in both real and virtual worlds, large amount of data and information will flow between co-spaces, which pose new challenges. For example, it would be challenging to process heterogeneous data streams in order to model and simulate real world events in the virtual world. Besides, more intelligent processing is needed to identify and send interesting events in the co-space to objects in the physical world.
      
  \item {\it Transaction Handling:} When the data being updated is spread across hundreds or thousands of networked computers/smart things with differing update policies, it would be difficult to define what the transaction is. In addition, most of things are resource-constrained, which are typically connected to the Internet using light-weight, {\em stateless} protocols such as CoAP (Constraint Application Protocol)\footnote{http://tools.ietf.org/html/draft-ietf-core-coap-18} and 6LoWPAN (IPv6 over Low Power Wireless Personal Area Networks)\footnote{http://tools.ietf.org/wg/6lowpan} and accessed using RESTful Web services.
This makes transaction handling in IoT a great challenge. As \citeN{JamesCJS09} point out that the problem is that the world is changing fast, the data representing the world is on multiple networked computers/smart things and existing database technologies cannot manage. Techniques developed for streamed and real-time data may provide some hints.

  \item {\it Frequently Updated Timestamped Structured (FUTS) Data:} The Internet, and hence IoT, contains potentially billions of Frequently Updated Timestamped Structured (FUTS) data sources, such as real-time traffic reports, air pollution detection, temperature monitoring, crops monitoring, etc. FUTS data sources contain states and updates of  physical world things. Current technologies are not capable in dealing with FUTS data sources \cite{JamesCJS09} because: (i) no data management system can easily display FUTS past data; (ii) no efficient crawler or storage engine is able to collect and store FUTS data; and (iii) querying and delivering FUTS data is hardly supported. All these pose great challenges for the design of novel data management systems for FUTS data.

  \item {\it Distributed and Mobile Data:} In IoT, data will be increasingly distributed and mobile \cite{JamesCJS09}. 
Different from traditional mobile data, distributed and mobile data in IoT would be much more highly distributed and data intensive. In the context of interconnecting huge numbers of mobile and smart objects, centralized data stores would not be a suitable tool to manage all the dynamics of mobile data produced in IoT. Thus there is a need for novel ways to manage distributed and mobile data efficiently and effectively in IoT.
  
  \item {\it Semantic Enrichment and Semantic Event Processing:} The full potentials of IoT would heavily rely on the progress of semantic Web. This is because things and machines should play a much more important role than humans in IoT to process and understand data. This calls for new research in Semantic technologies. For example, there are increasing efforts in building public knowledge bases (such as DBpedia, FreeBase, Linked Open Data Cloud, etc.). But how these knowledge bases can be effectively used to add to the understanding of raw data coming from sensor data streams and other types of data streams? To resolve this challenge, semantic enrichment of sensing data is a promising research direction. Further, consider the potential excessively large amount of subscriptions of IoT data. To produce proper semantic enrichment to meet different enrichment needs from different subscribers poses great challenges. Finally, how to effectively incorporate semantic enrichment techniques with semantic event processing to provide much better expressiveness in event processing is still at its initial stage. This will also demand a large amount of research efforts.


  \item {\it Mining:} Data mining aims to facilitate the exploration and analysis of large amounts of data, which can help to extract useful information for huge volume of IoT data. Data mining challenges may include extraction of temporal characteristics from sensor data streams, event detection from multiple data streams, data stream classification, activity discovery and recognition from sensor data streams. Besides, clustering and table summarization in large data sets, mining large (data, information or social) networks, sampling, and information extraction from the Web are also great challenges in IoT.

  \item {\it Knowledge Discovery:} Knowledge discovery is the process of extracting useful knowledge from data. This is essential especially when connected things populate their data to the Web. \citeN{Weikum11} identify the following issues related to knowledge discovery in IoT: (i) automatic extraction of relational facts from natural-language text and multi-modal contexts; (ii) large-scale gathering of factual-knowledge candidates and their reconciliation into comprehensive knowledge bases; (iii) reasoning on uncertain hypotheses, for knowledge discovery and semantic search; and (iv) deep and real-time question answering, e.g., to enable computers to win quiz game shows.



  \item {\it Security:} Due to the proliferation of embedded devices in IoT, effective device security mechanisms are essential to the development of IoT technologies and applications. National Intelligence Council \cite{NICDCT08} argues that, to the extent that everyday objects become information security risks, the IoT could distribute those risks far more widely than the Internet has to date. For example, RFID security presents many challenges. Potential solutions should consider aspects from hardware and wireless protocol security to the management, regulation and sharing of collected RFID data \cite{WelbourneBCGRRBB09}. \citeN{LinSHMWKLPSY09} point out that establishing trust between a group of individuals remains a difficult problem. Besides, \citeN{LagesseKPW09} argue that there is still no generic framework for deploying and extending traditional security mechanisms over a variety of pervasive systems. Regarding security concerns of the network layer, \citeN{KounavisKGEGD10} suggest that the Internet can be gradually encrypted and authenticated based on the observations that the recent advances in implementation of cryptographic algorithms have made general purpose processors capable of encrypting packets at high rates. But how to generalize such algorithms to IoT would be challenging as things in IoT normally only maintain low transmission rates and connections are usually intermittent.

  \item {\it Privacy:} Privacy protection is a serious challenge in IoT. One of the fundamental problems is the lack of a mechanism to help people expose appropriate amounts of their identity information \cite{ZhuZ09}.  Embedded sensing is becoming more and more prevalent on personal devices such as mobile phones and multi-media players. Since people are typically wearing and carrying devices capable of sensing,  details such as activity, location, and environment could become available to other people. Hence, personal sensing can be used to detect their physical activities and bring about privacy concerns \cite{KlasnjaCCBH09}. 

  \item {\it Social Concerns:} Since IoT connects everyday objects to the Internet, social concerns would become a hot topic in the development of IoT. For example, home is a private and intimate place. It may have multiple stakeholders competing priorities and tolerances for what is acceptable and useful \cite{ChoeCJHK11}. To build smart homes in IoT, similar social concerns should be considered. Further, online social networks with personal things information may incur social concerns as well, such as disclosures of personal activities and hobbies, etc. Appropriate economic and legal conditions and a social consensus on how the new technical opportunities in IoT should be used also represents a substantial task for the future \cite{MatternF10}.
\end{itemize}

\section{Summary}\label{sec:summary}
It is widely predicted that the next generation of the Internet will be comprised of trillions of connected computing nodes at a global scale. Through these nodes, everyday objects in the world can be identified, connected to the Internet and take decisions independently. In this context, Internet of Things (IoT) is considered a new revolution of the Internet. In IoT, the possibility of seamlessly merging the real and the virtual worlds, through the massive deployment of embedded devices, opens up many new and exciting directions for both research and development. In this article, we have provided an overview of some key research areas of IoT, specifically from a data-centric perspective. It also presents a number of fundamental issues to be resolved before we can fully realize the promise of IoT applications. This article covers investigations on data models, data storage, stream processing, search and event processing. The most relevant application fields have also been reviewed. 

Over the last few years, the Internet of Things has gained momentum and is becoming a rapidly expanding area of research and business. Many efforts from researchers, vendors and governments have been devoted to creating and developing novel IoT applications. Along with the current research efforts, we encourage more insights into the problems of this promising technology, and more efforts in addressing the open research issues identified in this article. 

\bibliographystyle{acmtrans}
\bibliography{survey}

\begin{thebibliography}{}

\bibitem[\protect\citeauthoryear{Abadi, Marcus, Madden, and Hollenbach}{Abadi
  et~al\mbox{.}}{2007}]{AbadiMMH07}
{\sc Abadi, D.~J.}, {\sc Marcus, A.}, {\sc Madden, S.}, {\sc and} {\sc
  Hollenbach, K.~J.} 2007.
\newblock {Scalable Semantic Web Data Management Using Vertical Partitioning}.
\newblock In {\em Proceedings of the 33rd International Conference on Very
  Large Data Bases (VLDB)}. ACM, University of Vienna, Austria, 411--422.

\bibitem[\protect\citeauthoryear{Agrawal, Diao, Gyllstrom, and
  Immerman}{Agrawal et~al\mbox{.}}{2008}]{sigmod/AgrawalDGI08}
{\sc Agrawal, J.}, {\sc Diao, Y.}, {\sc Gyllstrom, D.}, {\sc and} {\sc
  Immerman, N.} 2008.
\newblock {Efficient pattern matching over event streams}.
\newblock In {\em Proceedings of the ACM SIGMOD International Conference on
  Management of Data (SIGMOD'08)}. 147--160.

\bibitem[\protect\citeauthoryear{Agrawal, Chakrabarti, Chaudhuri, Ganti,
  K{\"o}nig, and Xin}{Agrawal et~al\mbox{.}}{2009}]{AgrawalCCGKX09}
{\sc Agrawal, S.}, {\sc Chakrabarti, K.}, {\sc Chaudhuri, S.}, {\sc Ganti, V.},
  {\sc K{\"o}nig, A.~C.}, {\sc and} {\sc Xin, D.} 2009.
\newblock {Exploiting Web Search Engines to Search Structured Databases}.
\newblock In {\em Proceedings of the 18th International Conference on World
  Wide Web (WWW)}. ACM, Madrid, Spain, 501--510.

\bibitem[\protect\citeauthoryear{Anicic, Fodor, Rudolph, and Stojanovic}{Anicic
  et~al\mbox{.}}{2011}]{www/AnicicFRS11}
{\sc Anicic, D.}, {\sc Fodor, P.}, {\sc Rudolph, S.}, {\sc and} {\sc
  Stojanovic, N.} 2011.
\newblock {EP-SPARQL: a unified language for event processing and stream
  reasoning}.
\newblock In {\em Proceedings of the 20th International Conference on World
  Wide Web (WWW)}. 635--644.

\bibitem[\protect\citeauthoryear{Anonymous}{Anonymous}{2008}]{NICDCT08}
{\sc Anonymous}. 2008.
\newblock {National Intelligence Council (NIC), Disruptive Civil Technologies:
  Six Technologies with Potential Impacts on US Interests Out to 2025,
  Conference Report CR 2008-07, April 2008,
  http://www.dni.gov/nic/NIC\_home.html}.

\bibitem[\protect\citeauthoryear{Apache}{Apache}{2014}]{HBase}
{\sc Apache}. 2014.
\newblock {Apache HBase Project, https://hbase.apache.org/}.

\bibitem[\protect\citeauthoryear{Artikis and Paliouras}{Artikis and
  Paliouras}{2014}]{edbt/ArtikisP14}
{\sc Artikis, A.} {\sc and} {\sc Paliouras, G.} 2014.
\newblock {Tutorial: Formal Methods for Event Processing}.
\newblock In {\em Proc. 17th International Conference on Extending Database
  Technology (EDBT)}. 675.

\bibitem[\protect\citeauthoryear{Ashton}{Ashton}{2009}]{AshtonK09}
{\sc Ashton, K.} 2009.
\newblock {That `Internet of Things' Thing.
  http://www.rfidjournal.com/article/view/4986}.

\bibitem[\protect\citeauthoryear{Atzori, Iera, and Morabito}{Atzori
  et~al\mbox{.}}{2010}]{AtzoriIM10}
{\sc Atzori, L.}, {\sc Iera, A.}, {\sc and} {\sc Morabito, G.} 2010.
\newblock {The Internet of Things: A survey}.
\newblock {\em Computer Networks\/}~{\em 54,\/}~15, 2787--2805.

\bibitem[\protect\citeauthoryear{Babcock and Olston}{Babcock and
  Olston}{2003}]{sigmod/BabcockO03}
{\sc Babcock, B.} {\sc and} {\sc Olston, C.} 2003.
\newblock {Distributed Top-K Monitoring}.
\newblock In {\em Proceedings of the 2003 ACM SIGMOD International Conference
  on Management of Data (SIGMOD Conference)}. 28--39.

\bibitem[\protect\citeauthoryear{Barbieri, Braga, Ceri, and
  Grossniklaus}{Barbieri et~al\mbox{.}}{2010}]{edbt/BarbieriBCG10}
{\sc Barbieri, D.~F.}, {\sc Braga, D.}, {\sc Ceri, S.}, {\sc and} {\sc
  Grossniklaus, M.} 2010.
\newblock {An execution environment for C-SPARQL queries}.
\newblock In {\em Proceedings of the 13th International Conference on Extending
  Database Technology (EDBT)}. 441--452.

\bibitem[\protect\citeauthoryear{Barnaghi, Sheth, and Henson}{Barnaghi
  et~al\mbox{.}}{2013}]{barnagi-is2013}
{\sc Barnaghi, P.~M.}, {\sc Sheth, A.~P.}, {\sc and} {\sc Henson, C.~A.} 2013.
\newblock {From Data to Actionable Knowledge: Big Data Challenges in the Web of
  Things}.
\newblock {\em IEEE Intelligent Systems\/}~{\em 28,\/}~6, 6--11.

\bibitem[\protect\citeauthoryear{Biddlecombe}{Biddlecombe}{2005}]{Biddlecombe05}
{\sc Biddlecombe, E.} 2005.
\newblock {UN predicts 'Internet of Things'.
  http://news.bbc.co.uk/2/hi/technology/4440334.stm}.

\bibitem[\protect\citeauthoryear{Bolles, Grawunder, and Jacobi}{Bolles
  et~al\mbox{.}}{2008}]{esws/BollesGJ08}
{\sc Bolles, A.}, {\sc Grawunder, M.}, {\sc and} {\sc Jacobi, J.} 2008.
\newblock {Streaming SPARQL - Extending SPARQL to Process Data Streams}.
\newblock In {\em Proceedings of the 5th European Semantic Web Conference on
  The Semantic Web: Research and Applications (ESWC)}. 448--462.

\bibitem[\protect\citeauthoryear{Bort}{Bort}{2011}]{Bort11}
{\sc Bort, J.} 2011.
\newblock {10 technologies that will change the world in the next 10 years.
  http://www.networkworld.com/news/2011/071511-cisco-futurist.html}.

\bibitem[\protect\citeauthoryear{Calbimonte, Corcho, and Gray}{Calbimonte
  et~al\mbox{.}}{2010}]{semweb/CalbimonteCG10}
{\sc Calbimonte, J.-P.}, {\sc Corcho, {\'O}.}, {\sc and} {\sc Gray, A. J.~G.}
  2010.
\newblock {Enabling Ontology-Based Access to Streaming Data Sources}.
\newblock In {\em Proceedings of the 9th International Semantic Web Conference
  (ISWC)}. 96--111.

\bibitem[\protect\citeauthoryear{Cal\`{\i} and Martinenghi}{Cal\`{\i} and
  Martinenghi}{2010}]{CaliM10}
{\sc Cal\`{\i}, A.} {\sc and} {\sc Martinenghi, D.} 2010.
\newblock {Querying the Deep Web}.
\newblock In {\em Proceedings of the 13th International Conference on Extending
  Database Technology (EDBT)}. ACM, Lausanne, Switzerland, 724--727.

\bibitem[\protect\citeauthoryear{Cao, Sutton, Diao, and Shenoy}{Cao
  et~al\mbox{.}}{2011a}]{CaoSDS11}
{\sc Cao, Z.}, {\sc Sutton, C.}, {\sc Diao, Y.}, {\sc and} {\sc Shenoy, P.~J.}
  2011a.
\newblock {Distributed Inference and Query Processing for RFID Tracking and
  Monitoring}.
\newblock {\em Proceedings of the VLDB Endowment\/}~{\em 4,\/}~5, 326--337.

\bibitem[\protect\citeauthoryear{Cao, Sutton, Diao, and Shenoy}{Cao
  et~al\mbox{.}}{2011b}]{pvldb/CaoSDS11}
{\sc Cao, Z.}, {\sc Sutton, C.~A.}, {\sc Diao, Y.}, {\sc and} {\sc Shenoy,
  P.~J.} 2011b.
\newblock {Distributed inference and query processing for RFID tracking and
  monitoring}.
\newblock {\em Proceedings of the VLDB Endowment\/}~{\em 4,\/}~5, 326--337.

\bibitem[\protect\citeauthoryear{CASAGRAS}{CASAGRAS}{2000}]{CASAGRAS}
{\sc CASAGRAS}. 2000.
\newblock {CASAGRAS (Coordination And Support Action for Global RFID-related
  Activities and Standardisation).}

\bibitem[\protect\citeauthoryear{Chang, Dean, Ghemawat, Hsieh, Wallach,
  Burrows, Chandra, Fikes, and Gruber}{Chang
  et~al\mbox{.}}{2008}]{tocs/ChangDGHWBCFG08}
{\sc Chang, F.}, {\sc Dean, J.}, {\sc Ghemawat, S.}, {\sc Hsieh, W.~C.}, {\sc
  Wallach, D.~A.}, {\sc Burrows, M.}, {\sc Chandra, T.}, {\sc Fikes, A.}, {\sc
  and} {\sc Gruber, R.~E.} 2008.
\newblock {Bigtable: A Distributed Storage System for Structured Data}.
\newblock {\em ACM Trans. Comput. Syst.\/}~{\em 26,\/}~2.

\bibitem[\protect\citeauthoryear{Chaudhuri, Ganti, and Xin}{Chaudhuri
  et~al\mbox{.}}{2009}]{ChaudhuriGX09}
{\sc Chaudhuri, S.}, {\sc Ganti, V.}, {\sc and} {\sc Xin, D.} 2009.
\newblock {Exploiting Web Search to Generate Synonyms for Entities}.
\newblock In {\em Proceedings of the 18th International Conference on World
  Wide Web (WWW)}. ACM, Madrid, Spain, 151--160.

\bibitem[\protect\citeauthoryear{Chen, Li, Ooi, and Wu}{Chen
  et~al\mbox{.}}{2011}]{ChenLOW11}
{\sc Chen, C.}, {\sc Li, F.}, {\sc Ooi, B.~C.}, {\sc and} {\sc Wu, S.} 2011.
\newblock {TI: An Efficient Indexing Mechanism for Real-Time Search on Tweets}.
\newblock In {\em Proceedings of the ACM SIGMOD International Conference on
  Management of Data (SIGMOD)}. ACM, Athens, Greece, 649--660.

\bibitem[\protect\citeauthoryear{Chen, Mao, and Liu}{Chen
  et~al\mbox{.}}{2014}]{monet/ChenML14}
{\sc Chen, M.}, {\sc Mao, S.}, {\sc and} {\sc Liu, Y.} 2014.
\newblock {Big Data: A Survey}.
\newblock {\em MONET\/}~{\em 19,\/}~2, 171--209.

\bibitem[\protect\citeauthoryear{Cho and Garcia-Molina}{Cho and
  Garcia-Molina}{2010}]{ChoG10}
{\sc Cho, J.} {\sc and} {\sc Garcia-Molina, H.} 2010.
\newblock {Dealing with Web Data: History and Look ahead}.
\newblock {\em Proceedings of the VLDB Endowment\/}~{\em 3,\/}~1, 4.

\bibitem[\protect\citeauthoryear{Choe, Consolvo, Jung, Harrison, and
  Kientz}{Choe et~al\mbox{.}}{2011}]{ChoeCJHK11}
{\sc Choe, E.~K.}, {\sc Consolvo, S.}, {\sc Jung, J.}, {\sc Harrison, B.~L.},
  {\sc and} {\sc Kientz, J.~A.} 2011.
\newblock {Living in a Glass House: A Survey of Private Moments in the Home}.
\newblock In {\em Proceedings of the 13th International Conference on
  Ubiquitous Computing (Ubicomp)}. ACM, Beijing, China, 41--44.

\bibitem[\protect\citeauthoryear{Christophe, Verdot, and Toubiana}{Christophe
  et~al\mbox{.}}{2011}]{ChristopheVT11}
{\sc Christophe, B.}, {\sc Verdot, V.}, {\sc and} {\sc Toubiana, V.} 2011.
\newblock {Searching the 'Web of Things'}.
\newblock In {\em Proceedings of the 5th IEEE International Conference on
  Semantic Computing (ICSC)}. IEEE, Palo Alto, CA, USA, 308--315.

\bibitem[\protect\citeauthoryear{Commission}{Commission}{2009}]{EUCommission09}
{\sc Commission, E.} 2009.
\newblock {European Commission: Internet of Things, An action plan for Europe.
  http://eur-lex.europa.eu/LexUriServ/site/en/com/2009/com2009\_0278en01.pdf}.

\bibitem[\protect\citeauthoryear{Cornelius and Kotz}{Cornelius and
  Kotz}{2011}]{CorneliusK11}
{\sc Cornelius, C.} {\sc and} {\sc Kotz, D.} 2011.
\newblock {Recognizing Whether Sensors Are on the Same Body}.
\newblock In {\em Proceedings of the 9th International Conference on Pervasive
  Computing (Pervasive)}. Springer, San Francisco, CA, USA, 332--349.

\bibitem[\protect\citeauthoryear{Cugola and Margara}{Cugola and
  Margara}{2012}]{csur/CugolaM12}
{\sc Cugola, G.} {\sc and} {\sc Margara, A.} 2012.
\newblock {Processing flows of information: From data stream to complex event
  processing}.
\newblock {\em ACM Comput. Surv.\/}~{\em 44,\/}~3, 15.

\bibitem[\protect\citeauthoryear{de~Andrade~Silva, Faria, Barros, Hruschka,
  de~Carvalho, and Gama}{de~Andrade~Silva
  et~al\mbox{.}}{2013}]{csur/SilvaFBHCG13}
{\sc de~Andrade~Silva, J.}, {\sc Faria, E.~R.}, {\sc Barros, R.~C.}, {\sc
  Hruschka, E.~R.}, {\sc de~Carvalho, A. C. P. L.~F.}, {\sc and} {\sc Gama, J.}
  2013.
\newblock {Data stream clustering: A survey}.
\newblock {\em ACM Comput. Surv.\/}~{\em 46,\/}~1, 13.

\bibitem[\protect\citeauthoryear{DeCandia, Hastorun, Jampani, Kakulapati,
  Lakshman, Pilchin, Sivasubramanian, Vosshall, and Vogels}{DeCandia
  et~al\mbox{.}}{2007}]{sosp/DeCandiaHJKLPSVV07}
{\sc DeCandia, G.}, {\sc Hastorun, D.}, {\sc Jampani, M.}, {\sc Kakulapati,
  G.}, {\sc Lakshman, A.}, {\sc Pilchin, A.}, {\sc Sivasubramanian, S.}, {\sc
  Vosshall, P.}, {\sc and} {\sc Vogels, W.} 2007.
\newblock {Dynamo: amazon's highly available key-value store}.
\newblock In {\em Proceedings of the 21st ACM Symposium on Operating Systems
  Principles (SOSP)}. 205--220.

\bibitem[\protect\citeauthoryear{Detweiler, Doniec, Jiang, Schwager, Chen, and
  Rus}{Detweiler et~al\mbox{.}}{2010}]{DetweilerDJSCR10}
{\sc Detweiler, C.}, {\sc Doniec, M.}, {\sc Jiang, M.}, {\sc Schwager, M.},
  {\sc Chen, R.}, {\sc and} {\sc Rus, D.} 2010.
\newblock {Adaptive Decentralized Control of Underwater Sensor Networks for
  Modeling Underwater Phenomena}.
\newblock In {\em Proceedings of the 8th International Conference on Embedded
  Networked Sensor Systems (SenSys)}. ACM, Zurich, Switzerland, 253--266.

\bibitem[\protect\citeauthoryear{Dong, Zhang, Kolari, Bai, Diaz, Chang, Zheng,
  and Zha}{Dong et~al\mbox{.}}{2010}]{DongZKBDCZZ10}
{\sc Dong, A.}, {\sc Zhang, R.}, {\sc Kolari, P.}, {\sc Bai, J.}, {\sc Diaz,
  F.}, {\sc Chang, Y.}, {\sc Zheng, Z.}, {\sc and} {\sc Zha, H.} 2010.
\newblock {Time is of the Essence: Improving Recency Ranking Using Twitter
  Data}.
\newblock In {\em Proceedings of the 19th International Conference on World
  Wide Web (WWW)}. ACM, Raleigh, North Carolina, USA, 331--340.

\bibitem[\protect\citeauthoryear{Elahi, R{\"o}mer, Ostermaier, Fahrmair, and
  Kellerer}{Elahi et~al\mbox{.}}{2009}]{ElahiROFK09}
{\sc Elahi, B.~M.}, {\sc R{\"o}mer, K.}, {\sc Ostermaier, B.}, {\sc Fahrmair,
  M.}, {\sc and} {\sc Kellerer, W.} 2009.
\newblock {Sensor Ranking: A Primitive for Efficient Content-based Sensor
  Search}.
\newblock In {\em Proceedings of the 8th International Conference on
  Information Processing in Sensor Networks (IPSN)}. IEEE, San Francisco,
  California, USA, 217--228.

\bibitem[\protect\citeauthoryear{Fan, Geerts, Ma, and M{\"u}ller}{Fan
  et~al\mbox{.}}{2010}]{FanGMM10}
{\sc Fan, W.}, {\sc Geerts, F.}, {\sc Ma, S.}, {\sc and} {\sc M{\"u}ller, H.}
  2010.
\newblock {Detecting Inconsistencies in Distributed Data}.
\newblock In {\em Proceedings of the 26th International Conference on Data
  Engineering (ICDE)}. IEEE, Long Beach, California, USA, 64--75.

\bibitem[\protect\citeauthoryear{Fazzinga, Flesca, Furfaro, and
  Parisi}{Fazzinga et~al\mbox{.}}{2014}]{edbt/FazzingaFFP14}
{\sc Fazzinga, B.}, {\sc Flesca, S.}, {\sc Furfaro, F.}, {\sc and} {\sc Parisi,
  F.} 2014.
\newblock {Cleaning trajectory data of RFID-monitored objects through
  conditioning under integrity constraints}.
\newblock In {\em Proc. 17th International Conference on Extending Database
  Technology (EDBT)}. 379--390.

\bibitem[\protect\citeauthoryear{Frank, Bolliger, Mattern, and Kellerer}{Frank
  et~al\mbox{.}}{2008}]{FrankBMK08}
{\sc Frank, C.}, {\sc Bolliger, P.}, {\sc Mattern, F.}, {\sc and} {\sc
  Kellerer, W.} 2008.
\newblock {The Sensor Internet at Work: Locating Everyday Items Using Mobile
  Phones}.
\newblock {\em Pervasive and Mobile Computing\/}~{\em 4,\/}~3, 421--447.

\bibitem[\protect\citeauthoryear{Gama}{Gama}{2010}]{books/daglib/0030859}
{\sc Gama, J.} 2010.
\newblock {\em {Knowledge Discovery from Data Streams}}.
\newblock Chapman and Hall / CRC Data Mining and Knowledge Discovery Series.
  CRC Press.

\bibitem[\protect\citeauthoryear{Ganti, Pham, Ahmadi, Nangia, and
  Abdelzaher}{Ganti et~al\mbox{.}}{2010}]{GantiPANA10}
{\sc Ganti, R.~K.}, {\sc Pham, N.}, {\sc Ahmadi, H.}, {\sc Nangia, S.}, {\sc
  and} {\sc Abdelzaher, T.~F.} 2010.
\newblock {GreenGPS: A Participatory Sensing Fuel-Efficient Maps Application}.
\newblock In {\em Proceedings of the 8th International Conference on Mobile
  Systems, Applications, and Services (MobiSys)}. ACM, San Francisco,
  California, USA, 151--164.

\bibitem[\protect\citeauthoryear{Gao, Guibas, Milosavljevic, and Zhou}{Gao
  et~al\mbox{.}}{2009}]{GaoGMZ09}
{\sc Gao, J.}, {\sc Guibas, L.~J.}, {\sc Milosavljevic, N.}, {\sc and} {\sc
  Zhou, D.} 2009.
\newblock {Distributed Resource Management and Matching in Sensor Networks}.
\newblock In {\em Proceedings of the 8th International Conference on
  Information Processing in Sensor Networks (IPSN)}. IEEE, San Francisco,
  California, USA, 97--108.

\bibitem[\protect\citeauthoryear{Garfinkel and Rosenberg}{Garfinkel and
  Rosenberg}{2005}]{GarfinkelR05}
{\sc Garfinkel, S.} {\sc and} {\sc Rosenberg, B.} 2005.
\newblock {\em {RFID: Applications, Security, and Privacy}}.
\newblock Addison-Wesley, Boston, USA.

\bibitem[\protect\citeauthoryear{Gerber, Hellmann, B{\"u}hmann, Soru, Usbeck,
  and Ngomo}{Gerber et~al\mbox{.}}{2013}]{semweb/GerberHBSUN13}
{\sc Gerber, D.}, {\sc Hellmann, S.}, {\sc B{\"u}hmann, L.}, {\sc Soru, T.},
  {\sc Usbeck, R.}, {\sc and} {\sc Ngomo, A.-C.~N.} 2013.
\newblock {Real-Time RDF Extraction from Unstructured Data Streams}.
\newblock In {\em Proceedings of the 12th International Semantic Web Conference
  (ISWC)}. 135--150.

\bibitem[\protect\citeauthoryear{Gilbert and Lynch}{Gilbert and
  Lynch}{2002}]{sigact/GilbertL02}
{\sc Gilbert, S.} {\sc and} {\sc Lynch, N.~A.} 2002.
\newblock {Brewer's conjecture and the feasibility of consistent, available,
  partition-tolerant web services}.
\newblock {\em SIGACT News\/}~{\em 33,\/}~2, 51--59.

\bibitem[\protect\citeauthoryear{Guha, Plarre, Lissner, Mitra, Krishna, Dutta,
  and Kumar}{Guha et~al\mbox{.}}{2010}]{GuhaPLMKDK10}
{\sc Guha, S.}, {\sc Plarre, K.}, {\sc Lissner, D.}, {\sc Mitra, S.}, {\sc
  Krishna, B.}, {\sc Dutta, P.}, {\sc and} {\sc Kumar, S.} 2010.
\newblock {AutoWitness: Locating and Tracking Stolen Property While Tolerating
  GPS and Radio Outages}.
\newblock In {\em Proceedings of the 8th International Conference on Embedded
  Networked Sensor Systems (SenSys)}. ACM, Zurich, Switzerland, 29--42.

\bibitem[\protect\citeauthoryear{Guinard}{Guinard}{2010}]{Guinard10}
{\sc Guinard, D.} 2010.
\newblock {A Web of Things for Smarter Cities}.
\newblock In {\em Technical Talk}. 1--8.

\bibitem[\protect\citeauthoryear{Guo, Zhang, Tan, and Guo}{Guo
  et~al\mbox{.}}{2011}]{cikm/GuoZTG11}
{\sc Guo, J.}, {\sc Zhang, P.}, {\sc Tan, J.}, {\sc and} {\sc Guo, L.} 2011.
\newblock {Mining frequent patterns across multiple data streams}.
\newblock In {\em Proceedings of the 20th ACM Conference on Information and
  Knowledge Management (CIKM)}. 2325--2328.

\bibitem[\protect\citeauthoryear{Gupta, Intille, and Larson}{Gupta
  et~al\mbox{.}}{2009}]{GuptaIL09}
{\sc Gupta, M.}, {\sc Intille, S.~S.}, {\sc and} {\sc Larson, K.} 2009.
\newblock {Adding GPS-Control to Traditional Thermostats: An Exploration of
  Potential Energy Savings and Design Challenges}.
\newblock In {\em Proceedings of the 7th International Conference on Pervasive
  Computing (Pervasive)}. Springer, Nara, Japan, 95--114.

\bibitem[\protect\citeauthoryear{Hasan, O'Riain, and Curry}{Hasan
  et~al\mbox{.}}{2012}]{debs/HasanOC12}
{\sc Hasan, S.}, {\sc O'Riain, S.}, {\sc and} {\sc Curry, E.} 2012.
\newblock {Approximate semantic matching of heterogeneous events}.
\newblock In {\em Proceedings of the Sixth ACM International Conference on
  Distributed Event-Based Systems (DEBS)}. 252--263.

\bibitem[\protect\citeauthoryear{Hasan, O'Riain, and Curry}{Hasan
  et~al\mbox{.}}{2013}]{debs/HasanOC13}
{\sc Hasan, S.}, {\sc O'Riain, S.}, {\sc and} {\sc Curry, E.} 2013.
\newblock {Towards unified and native enrichment in event processing systems}.
\newblock In {\em Proceedings of the 7th ACM International Conference on
  Distributed Event-Based Systems (DEBS)}. 171--182.

\bibitem[\protect\citeauthoryear{He, Barman, and Naughton}{He
  et~al\mbox{.}}{2014}]{icdt/HeBN14}
{\sc He, Y.}, {\sc Barman, S.}, {\sc and} {\sc Naughton, J.~F.} 2014.
\newblock {On Load Shedding in Complex Event Processing}.
\newblock In {\em Proc. 17th International Conference on Database Theory
  (ICDT)}. 213--224.

\bibitem[\protect\citeauthoryear{Heinze, Ji, Pan, Gr{\"u}neberger, Jerzak, and
  Fetzer}{Heinze et~al\mbox{.}}{2013}]{vldb/HeinzeJPGJF13}
{\sc Heinze, T.}, {\sc Ji, Y.}, {\sc Pan, Y.}, {\sc Gr{\"u}neberger, F.~J.},
  {\sc Jerzak, Z.}, {\sc and} {\sc Fetzer, C.} 2013.
\newblock Elastic complex event processing under varying query load.
\newblock In {\em Proceedings of the First International Workshop on Big
  Dynamic Distributed Data}. 25--30.

\bibitem[\protect\citeauthoryear{Huang, Wang, Jia, and Fuxman}{Huang
  et~al\mbox{.}}{2011}]{HuangWJF11}
{\sc Huang, J.}, {\sc Wang, H.}, {\sc Jia, Y.}, {\sc and} {\sc Fuxman, A.}
  2011.
\newblock {Link-based Hidden Attribute Discovery for Objects on Web}.
\newblock In {\em Proceedings of the 14th International Conference on Extending
  Database Technology (EDBT)}. ACM, Uppsala, Sweden, 473--484.

\bibitem[\protect\citeauthoryear{INFSO}{INFSO}{2008}]{INFSO08}
{\sc INFSO}. 2008.
\newblock {INFSO D.4 Networked Enterprise \& RFID INFSO G.2 Micro \&
  Nanosystems, in: Co-operation with the Working Group RFID of the ETP EPOSS,
  Internet of Things in 2020, Roadmap for the Future, Version 1.1, 27 May
  2008.}

\bibitem[\protect\citeauthoryear{ITU}{ITU}{2005}]{ITU05}
{\sc ITU}. 2005.
\newblock {International Telecommunication Union (ITU) Internet Reports, The
  Internet of Things, November 2005.}

\bibitem[\protect\citeauthoryear{James, Cooper, Jeffery, and Saake}{James
  et~al\mbox{.}}{2009}]{JamesCJS09}
{\sc James, A.~E.}, {\sc Cooper, J.}, {\sc Jeffery, K.~G.}, {\sc and} {\sc
  Saake, G.} 2009.
\newblock {Research Directions in Database Architectures for the Internet of
  Things: A Communication of the First International Workshop on Database
  Architectures for the Internet of Things (DAIT 2009)}.
\newblock In {\em Proceedings of the 26th British National Conference on
  Databases (BNCOD)}. Springer, Birmingham, UK, 225--233.

\bibitem[\protect\citeauthoryear{Jeffery, Garofalakis, and Franklin}{Jeffery
  et~al\mbox{.}}{2006}]{vldb/JefferyGF06}
{\sc Jeffery, S.~R.}, {\sc Garofalakis, M.~N.}, {\sc and} {\sc Franklin, M.~J.}
  2006.
\newblock {Adaptive Cleaning for RFID Data Streams}.
\newblock In {\em Proceedings of the 32nd International Conference on Very
  Large Data Bases (VLDB)}. 163--174.

\bibitem[\protect\citeauthoryear{Jin, Zhang, and Das}{Jin
  et~al\mbox{.}}{2011}]{JinZD11}
{\sc Jin, X.}, {\sc Zhang, N.}, {\sc and} {\sc Das, G.} 2011.
\newblock {Attribute Domain Ddiscovery for Hidden Web Databases}.
\newblock In {\em Proceedings of the ACM SIGMOD International Conference on
  Management of Data (SIGMOD)}. ACM, Athens, Greece, 553--564.

\bibitem[\protect\citeauthoryear{Kadambi, Chen, Cooper, Lomax, Ramakrishnan,
  Silberstein, Tam, and Garcia-Molina}{Kadambi
  et~al\mbox{.}}{2011}]{KadambiCCLRSTG11}
{\sc Kadambi, S.}, {\sc Chen, J.}, {\sc Cooper, B.~F.}, {\sc Lomax, D.}, {\sc
  Ramakrishnan, R.}, {\sc Silberstein, A.}, {\sc Tam, E.}, {\sc and} {\sc
  Garcia-Molina, H.} 2011.
\newblock {Where in the World is My Data?}
\newblock {\em Proceedings of the VLDB Endowment\/}~{\em 4,\/}~11, 1040--1050.

\bibitem[\protect\citeauthoryear{Khare, An, and Song}{Khare
  et~al\mbox{.}}{2010}]{KhareAS10}
{\sc Khare, R.}, {\sc An, Y.}, {\sc and} {\sc Song, I.-Y.} 2010.
\newblock {Understanding Deep Web Search Interfaces: A Survey}.
\newblock {\em SIGMOD Record\/}~{\em 39,\/}~1, 33--40.

\bibitem[\protect\citeauthoryear{Klasnja, Consolvo, Choudhury, Beckwith, and
  Hightower}{Klasnja et~al\mbox{.}}{2009}]{KlasnjaCCBH09}
{\sc Klasnja, P.~V.}, {\sc Consolvo, S.}, {\sc Choudhury, T.}, {\sc Beckwith,
  R.}, {\sc and} {\sc Hightower, J.} 2009.
\newblock {Exploring Privacy Concerns about Personal Sensing}.
\newblock In {\em Proceedings of the 7th International Conference on Pervasive
  Computing (Pervasive)}. Springer, Nara, Japan, 176--183.

\bibitem[\protect\citeauthoryear{Knorr and Ng}{Knorr and
  Ng}{1998}]{vldb/KnorrN98}
{\sc Knorr, E.~M.} {\sc and} {\sc Ng, R.~T.} 1998.
\newblock {Algorithms for Mining Distance-Based Outliers in Large Datasets}.
\newblock In {\em Proceedings of 24rd International Conference on Very Large
  Data Bases (VLDB)}. 392--403.

\bibitem[\protect\citeauthoryear{Koshizuka and Sakamura}{Koshizuka and
  Sakamura}{2010}]{KoshizukaS10}
{\sc Koshizuka, N.} {\sc and} {\sc Sakamura, K.} 2010.
\newblock {Ubiquitous ID: Standards for Ubiquitous Computing and the Internet
  of Things}.
\newblock {\em IEEE Pervasive Computing\/}~{\em 9,\/}~4, 98--101.

\bibitem[\protect\citeauthoryear{Kounavis, Kang, Grewal, Eszenyi, Gueron, and
  Durham}{Kounavis et~al\mbox{.}}{2010}]{KounavisKGEGD10}
{\sc Kounavis, M.~E.}, {\sc Kang, X.}, {\sc Grewal, K.}, {\sc Eszenyi, M.},
  {\sc Gueron, S.}, {\sc and} {\sc Durham, D.} 2010.
\newblock {Encrypting the Tnternet}.
\newblock In {\em Proceedings of the ACM SIGCOMM Conference on Applications,
  Technologies, Architectures, and Protocols for Computer Communications
  (SIGCOMM)}. ACM, New Delhi, India, 135--146.

\bibitem[\protect\citeauthoryear{Lagesse, Kumar, Paluska, and Wright}{Lagesse
  et~al\mbox{.}}{2009}]{LagesseKPW09}
{\sc Lagesse, B.}, {\sc Kumar, M.}, {\sc Paluska, J.~M.}, {\sc and} {\sc
  Wright, M.} 2009.
\newblock {DTT: A Distributed Trust Toolkit for Pervasive Systems}.
\newblock In {\em Proceedings of the 7th Annual IEEE International Conference
  on Pervasive Computing and Communications (PerCom)}. IEEE, Galveston, TX,
  USA, 1--8.

\bibitem[\protect\citeauthoryear{Lai, Chen, Huang, and Chu}{Lai
  et~al\mbox{.}}{2010}]{LaiCHC10}
{\sc Lai, T.-T.}, {\sc Chen, Y.-H.}, {\sc Huang, P.}, {\sc and} {\sc Chu,
  H.-H.} 2010.
\newblock {PipeProbe: A Mobile Sensor Droplet for Mapping Hidden Pipeline}.
\newblock In {\em Proceedings of the 8th International Conference on Embedded
  Networked Sensor Systems (SenSys)}. ACM, Zurich, Switzerland, 113--126.

\bibitem[\protect\citeauthoryear{Lakshman and Malik}{Lakshman and
  Malik}{2010}]{sigops/LakshmanM10}
{\sc Lakshman, A.} {\sc and} {\sc Malik, P.} 2010.
\newblock {Cassandra: a decentralized structured storage system}.
\newblock {\em Operating Systems Review\/}~{\em 44,\/}~2, 35--40.

\bibitem[\protect\citeauthoryear{Lester, Hartung, Pina, Libby, Borriello, and
  Duncan}{Lester et~al\mbox{.}}{2009}]{LesterHPLBD09}
{\sc Lester, J.}, {\sc Hartung, C.}, {\sc Pina, L.}, {\sc Libby, R.}, {\sc
  Borriello, G.}, {\sc and} {\sc Duncan, G.} 2009.
\newblock {Validated Caloric Expenditure Estimation Using a Single Body-Worn
  Sensor}.
\newblock In {\em Proceedings of the 11th International Conference on
  Ubiquitous Computing (Ubicomp)}. ACM, Orlando, Florida, USA, 225--234.

\bibitem[\protect\citeauthoryear{Li, Xu, and Zhao}{Li
  et~al\mbox{.}}{2014}]{Li-isf2014}
{\sc Li, S.}, {\sc Xu, L.~D.}, {\sc and} {\sc Zhao, S.} 2014.
\newblock {The Internet of Things: A Survey}.
\newblock {\em Information Systems Frontiers\/}, (to appear).

\bibitem[\protect\citeauthoryear{Lian and Chen}{Lian and Chen}{2011}]{LianC11}
{\sc Lian, X.} {\sc and} {\sc Chen, L.} 2011.
\newblock {Efficient Query Answering in Probabilistic RDF Graphs}.
\newblock In {\em Proceedings of the ACM SIGMOD International Conference on
  Management of Data (SIGMOD)}. ACM, Athens, Greece, 157--168.

\bibitem[\protect\citeauthoryear{Liao, Li, Chen, and Wan}{Liao
  et~al\mbox{.}}{2011}]{cikm/LiaoLCW11}
{\sc Liao, G.}, {\sc Li, J.}, {\sc Chen, L.}, {\sc and} {\sc Wan, C.} 2011.
\newblock {KLEAP: an efficient cleaning method to remove cross-reads in RFID
  streams}.
\newblock In {\em Proceedings of the 20th ACM Conference on Information and
  Knowledge Management (CIKM)}. 2209--2212.

\bibitem[\protect\citeauthoryear{Lin, Studer, Hsiao, McCune, Wang, Krohn, Lin,
  Perrig, Sun, and Yang}{Lin et~al\mbox{.}}{2009}]{LinSHMWKLPSY09}
{\sc Lin, Y.-H.}, {\sc Studer, A.}, {\sc Hsiao, H.-C.}, {\sc McCune, J.~M.},
  {\sc Wang, K.-H.}, {\sc Krohn, M.~N.}, {\sc Lin, P.-L.}, {\sc Perrig, A.},
  {\sc Sun, H.-M.}, {\sc and} {\sc Yang, B.-Y.} 2009.
\newblock {SPATE: Small-group PKI-less Authenticated Trust Establishment}.
\newblock In {\em Proceedings of the 7th International Conference on Mobile
  Systems, Applications, and Services (MobiSys)}. ACM, Krak{\'o}w, Poland,
  1--14.

\bibitem[\protect\citeauthoryear{Liu, Bunn, and Chandy}{Liu
  et~al\mbox{.}}{2011}]{LiuBC11}
{\sc Liu, A.~H.}, {\sc Bunn, J.~J.}, {\sc and} {\sc Chandy, K.~M.} 2011.
\newblock {Sensor Networks for the Detection and Tracking of Radiation and
  Other Threats in Cities}.
\newblock In {\em Proceedings of the 10th International Conference on
  Information Processing in Sensor Networks (IPSN)}. IEEE, Chicago, IL, USA,
  1--12.

\bibitem[\protect\citeauthoryear{Liu, Rundensteiner, Dougherty, Gupta, Wang,
  Ari, and Mehta}{Liu et~al\mbox{.}}{2011}]{icde/LiuRDGWAM11}
{\sc Liu, M.}, {\sc Rundensteiner, E.~A.}, {\sc Dougherty, D.~J.}, {\sc Gupta,
  C.}, {\sc Wang, S.}, {\sc Ari, I.}, {\sc and} {\sc Mehta, A.} 2011.
\newblock {High-performance nested CEP query processing over event streams}.
\newblock In {\em Proceedings of the 27th International Conference on Data
  Engineering (ICDE)}. 123--134.

\bibitem[\protect\citeauthoryear{Lu, Sookoor, Srinivasan, Gao, Holben,
  Stankovic, Field, and Whitehouse}{Lu et~al\mbox{.}}{2010}]{LuSSGHSFW10}
{\sc Lu, J.}, {\sc Sookoor, T.~I.}, {\sc Srinivasan, V.}, {\sc Gao, G.}, {\sc
  Holben, B.}, {\sc Stankovic, J.~A.}, {\sc Field, E.}, {\sc and} {\sc
  Whitehouse, K.} 2010.
\newblock {The Smart Thermostat: Using Occupancy Sensors to Save Energy in
  Homes}.
\newblock In {\em Proceedings of the 8th International Conference on Embedded
  Networked Sensor Systems (SenSys)}. ACM, Zurich, Switzerland, 211--224.

\bibitem[\protect\citeauthoryear{Mathew, Atif, Sheng, and Maamar}{Mathew
  et~al\mbox{.}}{2014}]{puc/Sujith14}
{\sc Mathew, S.~S.}, {\sc Atif, Y.}, {\sc Sheng, Q.~Z.}, {\sc and} {\sc Maamar,
  Z.} 2014.
\newblock {Building Sustainable Parking Lots with the Web of Things}.
\newblock {\em Personal and Ubiquitious Computing\/}~{\em 18,\/}~4, 895--907.

\bibitem[\protect\citeauthoryear{Mathur, Jin, Kasturirangan, Chandrasekaran,
  Xue, Gruteser, and Trappe}{Mathur et~al\mbox{.}}{2010}]{MathurJKCXGT10}
{\sc Mathur, S.}, {\sc Jin, T.}, {\sc Kasturirangan, N.}, {\sc Chandrasekaran,
  J.}, {\sc Xue, W.}, {\sc Gruteser, M.}, {\sc and} {\sc Trappe, W.} 2010.
\newblock {ParkNet: Drive-by Sensing of Road-Side Parking Statistics}.
\newblock In {\em Proceedings of the 8th International Conference on Mobile
  Systems, Applications, and Services (MobiSys)}. ACM, San Francisco,
  California, USA, 123--136.

\bibitem[\protect\citeauthoryear{Mattern and Floerkemeier}{Mattern and
  Floerkemeier}{2010}]{MatternF10}
{\sc Mattern, F.} {\sc and} {\sc Floerkemeier, C.} 2010.
\newblock {From the Internet of Computers to the Internet of Things}.
\newblock In {\em From Active Data Management to Event-Based Systems and More}.
  Springer, 242--259.

\bibitem[\protect\citeauthoryear{Mottola}{Mottola}{2010}]{Mottola10}
{\sc Mottola, L.} 2010.
\newblock {Programming Storage-Centric Sensor Networks with Squirrel}.
\newblock In {\em Proceedings of the 9th International Conference on
  Information Processing in Sensor Networks (IPSN)}. IEEE, Stockholm, Sweden,
  1--12.

\bibitem[\protect\citeauthoryear{Nath}{Nath}{2009}]{Nath09}
{\sc Nath, S.} 2009.
\newblock {Energy Efficient Sensor Data Logging with Amnesic Flash Storage}.
\newblock In {\em Proceedings of the 8th International Conference on
  Information Processing in Sensor Networks (IPSN)}. IEEE, San Francisco,
  California, USA, 157--168.

\bibitem[\protect\citeauthoryear{Nie, Cocci, Cao, Diao, and Shenoy}{Nie
  et~al\mbox{.}}{2012}]{tkde/NieCCDS12}
{\sc Nie, Y.}, {\sc Cocci, R.}, {\sc Cao, Z.}, {\sc Diao, Y.}, {\sc and} {\sc
  Shenoy, P.~J.} 2012.
\newblock {SPIRE: Efficient Data Inference and Compression over RFID Streams}.
\newblock {\em {IEEE Trans. Knowl. Data Eng.}\/}~{\em 24,\/}~1, 141--155.

\bibitem[\protect\citeauthoryear{Ooi, Tan, and Tung}{Ooi
  et~al\mbox{.}}{2009}]{OoiTT09}
{\sc Ooi, B.~C.}, {\sc Tan, K.-L.}, {\sc and} {\sc Tung, A. K.~H.} 2009.
\newblock {Sense The Physical, Walkthrough The Virtual, Manage The Co
  (existing) Spaces: A Database Perspective}.
\newblock {\em SIGMOD Record\/}~{\em 38,\/}~3, 5--10.

\bibitem[\protect\citeauthoryear{Ostermaier, R{\"o}mer, Mattern, Fahrmair, and
  Kellerer}{Ostermaier et~al\mbox{.}}{2010}]{OstermaierRMFK10}
{\sc Ostermaier, B.}, {\sc R{\"o}mer, K.}, {\sc Mattern, F.}, {\sc Fahrmair,
  M.}, {\sc and} {\sc Kellerer, W.} 2010.
\newblock {A Real-Time Search Engine for the Web of Things}.
\newblock In {\em Proceedings of the 2010 Internet of Things (IOT)}. IEEE,
  Tokyo, Japan.

\bibitem[\protect\citeauthoryear{Perera, Zaslavsky, Christen, and
  Georgakopoulos}{Perera et~al\mbox{.}}{2013}]{Perera-corr2013}
{\sc Perera, C.}, {\sc Zaslavsky, A.~B.}, {\sc Christen, P.}, {\sc and} {\sc
  Georgakopoulos, D.} 2013.
\newblock {Context Aware Computing for The Internet of Things: A Survey}.
\newblock {\em CoRR\/}~{\em abs/1305.0982}.

\bibitem[\protect\citeauthoryear{Phuoc, Dao-Tran, Parreira, and
  Hauswirth}{Phuoc et~al\mbox{.}}{2011}]{semweb/PhuocDPH11}
{\sc Phuoc, D.~L.}, {\sc Dao-Tran, M.}, {\sc Parreira, J.~X.}, {\sc and} {\sc
  Hauswirth, M.} 2011.
\newblock {A Native and Adaptive Approach for Unified Processing of Linked
  Streams and Linked Data}.
\newblock In {\em Proceedings of the 10th International Semantic Web Conference
  (ISWC)}. 370--388.

\bibitem[\protect\citeauthoryear{Prabhakar and Cheng}{Prabhakar and
  Cheng}{2009}]{PrabhakarC09}
{\sc Prabhakar, S.} {\sc and} {\sc Cheng, R.} 2009.
\newblock {Data Uncertainty Management in Sensor Networks}.
\newblock In {\em Encyclopedia of Database Systems}. Springer US, USA,
  647--651.

\bibitem[\protect\citeauthoryear{Priyantha, Kansal, Goraczko, and
  Zhao}{Priyantha et~al\mbox{.}}{2008}]{PriyanthaKGZ08}
{\sc Priyantha, N.~B.}, {\sc Kansal, A.}, {\sc Goraczko, M.}, {\sc and} {\sc
  Zhao, F.} 2008.
\newblock {Tiny Web Services: Design and Implementation of Interoperable and
  Evolvable Sensor Networks}.
\newblock In {\em Proceedings of the 6th International Conference on Embedded
  Networked Sensor Systems (SenSys)}. ACM, Raleigh, NC, USA, 253--266.

\bibitem[\protect\citeauthoryear{Ramanathan, Schoellhammer, Kohler, Whitehouse,
  Harmon, and Estrin}{Ramanathan et~al\mbox{.}}{2009}]{RamanathanSKWHE09}
{\sc Ramanathan, N.}, {\sc Schoellhammer, T.}, {\sc Kohler, E.}, {\sc
  Whitehouse, K.}, {\sc Harmon, T.}, {\sc and} {\sc Estrin, D.} 2009.
\newblock {Suelo: Human-Assisted Sensing for Exploratory Soil Monitoring
  Studies}.
\newblock In {\em Proceedings of the 7th International Conference on Embedded
  Networked Sensor Systems (SenSys)}. ACM, Berkeley, California, USA, 197--210.

\bibitem[\protect\citeauthoryear{Ray, Rundensteiner, Liu, Gupta, Wang, and
  Ari}{Ray et~al\mbox{.}}{2013}]{edbt/RayRLGWA13}
{\sc Ray, M.}, {\sc Rundensteiner, E.~A.}, {\sc Liu, M.}, {\sc Gupta, C.}, {\sc
  Wang, S.}, {\sc and} {\sc Ari, I.} 2013.
\newblock High-performance complex event processing using continuous sliding
  views.
\newblock In {\em Proceedings of Joint 2013 EDBT/ICDT Conferences (EDBT)}.
  525--536.

\bibitem[\protect\citeauthoryear{Sakaki, Okazaki, and Matsuo}{Sakaki
  et~al\mbox{.}}{2010}]{SakakiOM10}
{\sc Sakaki, T.}, {\sc Okazaki, M.}, {\sc and} {\sc Matsuo, Y.} 2010.
\newblock {Earthquake Shakes Twitter Users: Real-time Event Detection by Social
  Sensors}.
\newblock In {\em Proceedings of the 19th International Conference on World
  Wide Web (WWW)}. ACM, Raleigh, North Carolina, USA, 851--860.

\bibitem[\protect\citeauthoryear{Satpal, Bhadra, Sellamanickam, Rastogi, and
  Sen}{Satpal et~al\mbox{.}}{2011}]{SatpalBSRS11}
{\sc Satpal, S.}, {\sc Bhadra, S.}, {\sc Sellamanickam, S.}, {\sc Rastogi, R.},
  {\sc and} {\sc Sen, P.} 2011.
\newblock {Web Information Extraction Using Markov Logic Networks}.
\newblock In {\em Proceedings of the 17th ACM SIGKDD International Conference
  on Knowledge Discovery and Data Mining (KDD)}. ACM, San Diego, CA, USA,
  1406--1414.

\bibitem[\protect\citeauthoryear{Scott, Brush, Krumm, Meyers, Hazas, Hodges,
  and Villar}{Scott et~al\mbox{.}}{2011}]{ScottBKMHHV11}
{\sc Scott, J.}, {\sc Brush, A. J.~B.}, {\sc Krumm, J.}, {\sc Meyers, B.}, {\sc
  Hazas, M.}, {\sc Hodges, S.}, {\sc and} {\sc Villar, N.} 2011.
\newblock {PreHeat: Controlling Home Heating Using Occupancy Prediction}.
\newblock In {\em Proceedings of the 13th International Conference on
  Ubiquitous Computing (Ubicomp)}. ACM, Beijing, China, 281--290.

\bibitem[\protect\citeauthoryear{Sequeda and Corcho}{Sequeda and
  Corcho}{2009}]{semweb/SequedaC09}
{\sc Sequeda, J.} {\sc and} {\sc Corcho, {\'O}.} 2009.
\newblock {Linked Stream Data: A Position Paper}.
\newblock In {\em Proceedings of the 2nd International Workshop on Semantic
  Sensor Networks (SSN09)}. 148--157.

\bibitem[\protect\citeauthoryear{Sheng, Li, and Zeadally}{Sheng
  et~al\mbox{.}}{2008}]{Sheng-Computer2008}
{\sc Sheng, Q.~Z.}, {\sc Li, X.}, {\sc and} {\sc Zeadally, S.} 2008.
\newblock {Enabling Next-Generation RFID Applications: Solutions and
  Challenges}.
\newblock {\em IEEE Computer\/}~{\em 41,\/}~9 (September), 21--28.

\bibitem[\protect\citeauthoryear{Stephan, Meixner, Koessling, Floerchinger, and
  Ollinger}{Stephan et~al\mbox{.}}{2010}]{StephanMKFO10}
{\sc Stephan, P.}, {\sc Meixner, G.}, {\sc Koessling, H.}, {\sc Floerchinger,
  F.}, {\sc and} {\sc Ollinger, L.} 2010.
\newblock {Product-Mediated Communication through Digital Object Memories in
  Heterogeneous Value Chains}.
\newblock In {\em Proceedings of the 8th Annual IEEE International Conference
  on Pervasive Computing and Communications (PerCom)}. IEEE, Mannheim, Germany,
  199--207.

\bibitem[\protect\citeauthoryear{Stonebraker, Abadi, Batkin, Chen, Cherniack,
  Ferreira, Lau, Lin, Madden, O'Neil, O'Neil, Rasin, Tran, and
  Zdonik}{Stonebraker et~al\mbox{.}}{2005}]{StonebrakerABCCFLLMOORTZ05}
{\sc Stonebraker, M.}, {\sc Abadi, D.~J.}, {\sc Batkin, A.}, {\sc Chen, X.},
  {\sc Cherniack, M.}, {\sc Ferreira, M.}, {\sc Lau, E.}, {\sc Lin, A.}, {\sc
  Madden, S.}, {\sc O'Neil, E.~J.}, {\sc O'Neil, P.~E.}, {\sc Rasin, A.}, {\sc
  Tran, N.}, {\sc and} {\sc Zdonik, S.~B.} 2005.
\newblock {C-Store: A Column-oriented DBMS}.
\newblock In {\em Proceedings of the 31st International Conference on Very
  Large Data Bases (VLDB)}. ACM, Trondheim, Norway, 553--564.

\bibitem[\protect\citeauthoryear{Stonebraker, Madden, Abadi, Harizopoulos,
  Hachem, and Helland}{Stonebraker et~al\mbox{.}}{2007}]{StonebrakerMAHHH07}
{\sc Stonebraker, M.}, {\sc Madden, S.}, {\sc Abadi, D.~J.}, {\sc Harizopoulos,
  S.}, {\sc Hachem, N.}, {\sc and} {\sc Helland, P.} 2007.
\newblock {The End of an Architectural Era (It's Time for a Complete Rewrite)}.
\newblock In {\em Proceedings of the 33rd International Conference on Very
  Large Data Bases (VLDB)}. ACM, University of Vienna, Austria, 1150--1160.

\bibitem[\protect\citeauthoryear{Subramaniam and Gunopulos}{Subramaniam and
  Gunopulos}{2007}]{ads/SubramaniamG07}
{\sc Subramaniam, S.} {\sc and} {\sc Gunopulos, D.} 2007.
\newblock {A Survey of Stream Processing Problems and Techniques in Sensor
  Networks}.
\newblock In {\em Data Streams - Models and Algorithms}. 333--352.

\bibitem[\protect\citeauthoryear{Suchanek, Varde, Nayak, and
  Senellart}{Suchanek et~al\mbox{.}}{2011}]{SuchanekVNS11}
{\sc Suchanek, F.~M.}, {\sc Varde, A.~S.}, {\sc Nayak, R.}, {\sc and} {\sc
  Senellart, P.} 2011.
\newblock {The Hidden Web, XML and the Semantic Web: Scientific Data Management
  Perspectives}.
\newblock In {\em Proceedings of the 14th International Conference on Extending
  Database Technology (EDBT)}. ACM, Uppsala, Sweden, 534--537.

\bibitem[\protect\citeauthoryear{Tan, Sheng, Wang, and Li}{Tan
  et~al\mbox{.}}{2010}]{TanSWL10}
{\sc Tan, C.~C.}, {\sc Sheng, B.}, {\sc Wang, H.}, {\sc and} {\sc Li, Q.} 2010.
\newblock {Microsearch: A Search Engine for Embedded Devices Used in Pervasive
  Computing}.
\newblock {\em ACM Trans. Embedded Comput. Syst.\/}~{\em 9,\/}~4, 29.

\bibitem[\protect\citeauthoryear{Teymourian and Paschke}{Teymourian and
  Paschke}{2009}]{debs/TeymourianP09}
{\sc Teymourian, K.} {\sc and} {\sc Paschke, A.} 2009.
\newblock {Towards semantic event processing}.
\newblock In {\em Proceedings of the Third ACM International Conference on
  Distributed Event-Based Systems (DEBS)}.

\bibitem[\protect\citeauthoryear{Teymourian, Rohde, and Paschke}{Teymourian
  et~al\mbox{.}}{2012a}]{debs/TeymourianRP12}
{\sc Teymourian, K.}, {\sc Rohde, M.}, {\sc and} {\sc Paschke, A.} 2012a.
\newblock {Fusion of background knowledge and streams of events}.
\newblock In {\em DEBS}. 302--313.

\bibitem[\protect\citeauthoryear{Teymourian, Rohde, and Paschke}{Teymourian
  et~al\mbox{.}}{2012b}]{edbt/TeymourianRP12}
{\sc Teymourian, K.}, {\sc Rohde, M.}, {\sc and} {\sc Paschke, A.} 2012b.
\newblock {Knowledge-based processing of complex stock market events}.
\newblock In {\em Proceedings of the 15th International Conference on Extending
  Database Technology (EDBT)}. 594--597.

\bibitem[\protect\citeauthoryear{Tran, Sutton, Cocci, Nie, Diao, and
  Shenoy}{Tran et~al\mbox{.}}{2009}]{icde/TranSCNDS09}
{\sc Tran, T. T.~L.}, {\sc Sutton, C.~A.}, {\sc Cocci, R.}, {\sc Nie, Y.}, {\sc
  Diao, Y.}, {\sc and} {\sc Shenoy, P.~J.} 2009.
\newblock {Probabilistic Inference over RFID Streams in Mobile Environments}.
\newblock In {\em Proceedings of the 25th International Conference on Data
  Engineering (ICDE)}. 1096--1107.

\bibitem[\protect\citeauthoryear{Tsatsanifos, Sacharidis, and
  Sellis}{Tsatsanifos et~al\mbox{.}}{2011}]{TsatsanifosSS11}
{\sc Tsatsanifos, G.}, {\sc Sacharidis, D.}, {\sc and} {\sc Sellis, T.~K.}
  2011.
\newblock {On Enhancing Scalability for Distributed RDF/S Stores}.
\newblock In {\em Proceedings of the 14th International Conference on Extending
  Database Technology (EDBT)}. ACM, Uppsala, Sweden, 141--152.

\bibitem[\protect\citeauthoryear{Tsiftes and Dunkels}{Tsiftes and
  Dunkels}{2011}]{TsiftesD11}
{\sc Tsiftes, N.} {\sc and} {\sc Dunkels, A.} 2011.
\newblock {A Database in Every Sensor}.
\newblock In {\em Proceedings of the 9th International Conference on Embedded
  Networked Sensor Systems (SenSys)}. ACM, Seattle, WA, USA, 316--332.

\bibitem[\protect\citeauthoryear{Wang and Agrawal}{Wang and
  Agrawal}{2011}]{WangA11}
{\sc Wang, F.} {\sc and} {\sc Agrawal, G.} 2011.
\newblock {Effective and Efficient Sampling Methods for Deep Web Aggregation
  Queries}.
\newblock In {\em Proceedings of the 14th International Conference on Extending
  Database Technology (EDBT)}. ACM, Uppsala, Sweden, 425--436.

\bibitem[\protect\citeauthoryear{Wang and Liu}{Wang and
  Liu}{2011}]{comsur/WangL11}
{\sc Wang, F.} {\sc and} {\sc Liu, J.} 2011.
\newblock {Networked Wireless Sensor Data Collection: Issues, Challenges, and
  Approaches}.
\newblock {\em IEEE Communications Surveys and Tutorials\/}~{\em 13,\/}~4,
  673--687.

\bibitem[\protect\citeauthoryear{Wang, Tan, and Li}{Wang
  et~al\mbox{.}}{2010}]{WangTL10}
{\sc Wang, H.}, {\sc Tan, C.~C.}, {\sc and} {\sc Li, Q.} 2010.
\newblock {Snoogle: A Search Engine for Pervasive Environments}.
\newblock {\em IEEE Trans. Parallel Distrib. Syst.\/}~{\em 21,\/}~8,
  1188--1202.

\bibitem[\protect\citeauthoryear{Weikum}{Weikum}{2011}]{Weikum11}
{\sc Weikum, G.} 2011.
\newblock {Database Researchers: Plumbers or Thinkers?}
\newblock In {\em Proceedings of the 14th International Conference on Extending
  Database Technology (EDBT)}. ACM, Uppsala, Sweden, 9--10.

\bibitem[\protect\citeauthoryear{Welbourne, Battle, Cole, Gould, Rector,
  Raymer, Balazinska, and Borriello}{Welbourne
  et~al\mbox{.}}{2009}]{WelbourneBCGRRBB09}
{\sc Welbourne, E.}, {\sc Battle, L.}, {\sc Cole, G.}, {\sc Gould, K.}, {\sc
  Rector, K.}, {\sc Raymer, S.}, {\sc Balazinska, M.}, {\sc and} {\sc
  Borriello, G.} 2009.
\newblock {Building the Internet of Things Using RFID: The RFID Ecosystem
  Experience}.
\newblock {\em IEEE Internet Computing\/}~{\em 13,\/}~3, 48--55.

\bibitem[\protect\citeauthoryear{Wu, Diao, and Rizvi}{Wu
  et~al\mbox{.}}{2006}]{sigmod/WuDR06}
{\sc Wu, E.}, {\sc Diao, Y.}, {\sc and} {\sc Rizvi, S.} 2006.
\newblock {High-performance complex event processing over streams}.
\newblock In {\em Proceedings of the ACM SIGMOD International Conference on
  Management of Data (SIGMOD'06)}. 407--418.

\bibitem[\protect\citeauthoryear{Wu, Sheng, Shen, and Zeadally}{Wu
  et~al\mbox{.}}{2013}]{tpds/Wu13}
{\sc Wu, Y.}, {\sc Sheng, Q.~Z.}, {\sc Shen, H.}, {\sc and} {\sc Zeadally, S.}
  2013.
\newblock {Modeling Object Flows from Distributed and Federated RFID Data
  Streams for Efficient Tracking and Tracing}.
\newblock {\em IEEE Transactions on Parallel and Distributed Systems\/}~{\em
  24,\/}~10, 2036--2045.

\bibitem[\protect\citeauthoryear{Xie, Zhu, Sharaf, Zhou, and Pang}{Xie
  et~al\mbox{.}}{2012}]{cikm/XieZSzP12}
{\sc Xie, Q.}, {\sc Zhu, J.}, {\sc Sharaf, M.~A.}, {\sc Zhou, X.}, {\sc and}
  {\sc Pang, C.} 2012.
\newblock {Efficient buffer management for piecewise linear representation of
  multiple data streams}.
\newblock In {\em Proceedings of the 21st ACM International Conference on
  Information and Knowledge Management (CIKM)}. 2114--2118.

\bibitem[\protect\citeauthoryear{Yang, Wang, Noh, Le, and Abdelzaher}{Yang
  et~al\mbox{.}}{2009}]{YangWNLA09}
{\sc Yang, Y.}, {\sc Wang, L.}, {\sc Noh, D.~K.}, {\sc Le, H.~K.}, {\sc and}
  {\sc Abdelzaher, T.~F.} 2009.
\newblock {SolarStore: Enhancing Data Reliability in Solar-Powered
  Storage-Centric Sensor Networks}.
\newblock In {\em Proceedings of the 7th International Conference on Mobile
  Systems, Applications, and Services (MobiSys)}. ACM, Krak{\'o}w, Poland,
  333--346.

\bibitem[\protect\citeauthoryear{Yap, Srinivasan, and Motani}{Yap
  et~al\mbox{.}}{2008}]{YapSM08}
{\sc Yap, K.-K.}, {\sc Srinivasan, V.}, {\sc and} {\sc Motani, M.} 2008.
\newblock {MAX: Wide Area Human-Centric Search of the Physical World}.
\newblock {\em ACM Trans. on Sensor Networks\/}~{\em 4,\/}~4, 34.

\bibitem[\protect\citeauthoryear{Yuan, Zheng, Zhang, Xie, and Sun}{Yuan
  et~al\mbox{.}}{2011}]{YuanZZXS11}
{\sc Yuan, J.}, {\sc Zheng, Y.}, {\sc Zhang, L.}, {\sc Xie, X.}, {\sc and} {\sc
  Sun, G.} 2011.
\newblock {Where to Find My Next Passenger?}
\newblock In {\em Proceedings of the 13th International Conference on
  Ubiquitous Computing (Ubicomp)}. ACM, Beijing, China, 109--118.

\bibitem[\protect\citeauthoryear{Zeng, Guo, and Cheng}{Zeng
  et~al\mbox{.}}{2011}]{ZengGC11}
{\sc Zeng, D.}, {\sc Guo, S.}, {\sc and} {\sc Cheng, Z.} 2011.
\newblock {The Web of Things: A Survey (Invited Paper)}.
\newblock {\em Journal of Communications\/}~{\em 6,\/}~6, 424--438.

\bibitem[\protect\citeauthoryear{Zhang, Zhang, and Das}{Zhang
  et~al\mbox{.}}{2011}]{ZhangZD11}
{\sc Zhang, M.}, {\sc Zhang, N.}, {\sc and} {\sc Das, G.} 2011.
\newblock Mining a search engine's corpus: Efficient yet unbiased sampling and
  aggregate estimation.
\newblock In {\em Proceedings of the ACM SIGMOD International Conference on
  Management of Data (SIGMOD)}. ACM, Indianapolis, Indiana, USA, 793--804.

\bibitem[\protect\citeauthoryear{Zhang, Pham, Corcho, and Calbimonte}{Zhang
  et~al\mbox{.}}{2012}]{semweb/ZhangDCC12}
{\sc Zhang, Y.}, {\sc Pham, M.-D.}, {\sc Corcho, {\'O}.}, {\sc and} {\sc
  Calbimonte, J.-P.} 2012.
\newblock {SRBench: A Streaming RDF/SPARQL Benchmark}.
\newblock In {\em Proceedings of the 11th International Semantic Web Conference
  (ISWC)}. 641--657.

\bibitem[\protect\citeauthoryear{Zhou, Simmhan, and Prasanna}{Zhou
  et~al\mbox{.}}{2011}]{debs/ZhouSP11}
{\sc Zhou, Q.}, {\sc Simmhan, Y.}, {\sc and} {\sc Prasanna, V.~K.} 2011.
\newblock {Towards an inexact semantic complex event processing framework}.
\newblock In {\em Proceedings of the Fifth ACM International Conference on
  Distributed Event-Based Systems (DEBS)}. 401--402.

\bibitem[\protect\citeauthoryear{Zhu and Zhu}{Zhu and Zhu}{2009}]{ZhuZ09}
{\sc Zhu, F.} {\sc and} {\sc Zhu, W.} 2009.
\newblock {RationalExposure: a Game Theoretic Approach to Optimize Identity
  Exposure in Pervasive Computing Environments}.
\newblock In {\em Proceedings of the 7th Annual IEEE International Conference
  on Pervasive Computing and Communications (PerCom)}. IEEE, Galveston, TX,
  USA, 1--8.

\bibitem[\protect\citeauthoryear{Zimmerman and Robson}{Zimmerman and
  Robson}{2011}]{ZimmermanR11}
{\sc Zimmerman, T.} {\sc and} {\sc Robson, C.} 2011.
\newblock {Monitoring Residential Noise for Prospective Home Owners and
  Renters}.
\newblock In {\em Proceedings of the 9th International Conference on Pervasive
  Computing (Pervasive)}. Springer, San Francisco, CA, USA, 34--49.

\end{thebibliography}

\begin{received}
Received July 2014
\end{received}
\end{document}